\documentclass{rspublic}

\usepackage{amsfonts,float}
\usepackage{graphicx,epsfig}
\usepackage{psfrag}
\usepackage{cases}

\usepackage[english]{babel}
\usepackage[T1]{fontenc}
\usepackage[latin1]{inputenc} 
\usepackage{textcomp}
\usepackage{ae}
\usepackage{aecompl}
\usepackage[cm]{aeguill}
\usepackage{lmodern}
\usepackage{color}

\newtheorem{thrm}{Theorem}[section]

\newtheorem{rmrk}[thrm]{Remark}

\newcommand{\Q}{\ensuremath{\mathbb{Q}}}

\newcommand{\ms}{\overset{\,m}{\approx}}
\renewcommand{\cot}{{\mathcal C}}
\newcommand{\sit}{{\mathcal S}}
\renewcommand{\exp}[1]{e^{#1}}

\numberwithin{equation}{section}
\numberwithin{thrm}{section}
\everymath{\displaystyle}

\begin{document}

\title[A double-layer Boussinesq-type model]{A double-layer Boussinesq-type model for highly nonlinear and dispersive waves}
\author[F. Chazel, M. Benoit, A. Ern, S. Piperno]{F. Chazel$^{1,\thanks{Author for correspondence (florent-externe.chazel@edf.fr).}}$, M. Benoit$^1$, A. Ern$^2$ and S. Piperno$^2$}
\affiliation{1. Université Paris-Est, Saint-Venant Laboratory for Hydraulics\\(Joint research unit EDF R\&D - CETMEF - Ecole des Ponts)\\6 quai Watier, BP 49, F-78401 Chatou, France\\
\vspace{0.2em}2. Université Paris-Est, CERMICS, Ecole des Ponts\\6-8 avenue Blaise Pascal, F-77455 Marne la Vallée cedex 2, France}
\label{firstpage}
\maketitle

\begin{abstract}{Water waves; Boussinesq-type models; multi-layer models; velocity potential formulation; nonlinear dispersive waves; deep water; Padé approximants; velocity profiles; linear shoaling}
We derive and analyze in the framework of the mild-slope approximation a new double-layer Boussinesq-type model which is linearly and nonlinearly accurate up to deep water. Assuming the flow to be irrotational, we formulate the problem in terms of the velocity potential thereby lowering the number of unknowns. The model derivation combines two approaches, namely the method proposed by Agnon \textit{et al.} (Agnon \textit{et al.} 1999 \textit{J. Fluid Mech.} \textbf{399}, 319--333) and enhanced by Madsen \textit{et al.} (Madsen \textit{et al.} 2003 \textit{Proc. R. Soc. Lond. A} \textbf{459}, 1075--1104) which consists in constructing infinite-series Taylor solutions to the Laplace equation, to truncate them at a finite order and to use Padé approximants, and the double-layer approach of Lynett \& Liu (Lynett \& Liu 2004 \textit{Proc. R. Soc. Lond. A} \textbf{460}, 2637--2669) allowing to lower the order of derivatives. We formulate the model in terms of a static Dirichlet-Neumann operator translated from the free surface to the still-water level, and we derive an approximate inverse of this operator that can be built once and for all. The final model consists of only four equations both in one and two horizontal dimensions, and includes only second-order derivatives, which is a major improvement in comparison with so-called high-order Boussinesq models. A linear analysis of the model is performed and its properties are optimized using a free parameter determining the position of the interface between the two layers. Excellent dispersion and shoaling properties are obtained, allowing the model to be applied up to the deep-water value $kh=10$. Finally, numerical simulations are performed to quantify the nonlinear behaviour of the model, and the results exhibit a nonlinear range of validity reaching at least $kh=3\pi$. 
\end{abstract}

\section{Introduction}

During the past two decades, the original Boussinesq (1872) model for flat bottom and its uneven bottom version derived by Peregrine (1967) have been widely studied and extended to tackle more and more realistic physical problems. Consequently, Boussinesq-type models have emerged as an attractive and commonly used tool for coastal applications and engineering purposes. The derivation of such models is based on a polynomial approximation of the vertical profile of the velocity field, which allows to reduce the problem size from three to two space dimensions. These models are usually formulated as conservation equations for mass and momentum, including spatial and temporal derivatives of the free surface elevation and the velocity. In practice, their range of applicability is measured in terms of $kh$, where $k$ is the wavenumber and $h$ the water depth. 

The conventional Boussinesq model for uneven bottoms (Peregrine 1967), which employs a quadratic polynomial approximation for the vertical flow distribution, is a depth-averaged model based on two fundamental assumptions, namely weak nonlinearity and weak dispersion. Its range of applicability is limited to $kh < 0.75 $\textcolor{black}{, as stated in Madsen \textit{et al.} (2002, 2003),} so that this model has poor dispersion properties in intermediate depths. Moreover, the weakly nonlinear assumption limits the largest wave height that can be modelled accurately. As a result, substantial efforts have been devoted to extend the linear and nonlinear range of applicability of Boussinesq-type models. The first historical improvement, proposed by Nwogu (1993), consists in using a reference velocity at a specified depth, allowing the resulting model to be linearly applicable at intermediate depths. Similar models for short-amplitude and long waves have been more recently proposed and rigorously justified (Bona \textit{et al.} 2002, 2005; Chazel 2007). An effort similar to the one of Nwogu (1993) was pursued by Madsen \& S\o rensen (1992), which was followed by the works of Liu (1994) and Wei \textit{et al.} (1995) in which the authors efficiently removed the weak nonlinearity assumption, allowing the model to simulate wave propagation with strong nonlinear interaction. According to the reviews proposed by Madsen \& Schäffer (1998, 1999), these new Boussinesq-type models allow to extend the linear range of applicability to $kh = 6$, but it turned out that a similar improvement on the nonlinear characteristics was very difficult to reach. Then, so-called high-order Boussinesq-type models were derived to further enhance the deep-water linear and nonlinear accuracy using a higher-order (at least fourth-order) polynomial approximation for the vertical flow distribution. One such example is the formulation of Gobbi \textit{et al.} (2000) which uses a fourth-order polynomial approximation: this model exhibits excellent linear properties up to $kh = 6$ for the dispersion relation and up to $kh = 4$ for the vertical profiles of orbital velocities, whereas nonlinear behaviour is fairly well captured up to $kh = 3$. The price to pay for such an improvement is a significant increase in computational cost, since this model includes up to fifth-order derivatives and therefore requires a complex numerical scheme.

Over the past decade, two parallel approaches have emerged, one aiming at extending even more the range of applicability of the model of Gobbi \textit{et al.} (2000) into very deep water without increasing too much the numerical complexity, and another aiming at lowering the numerical cost of the latter while at least preserving the linear and nonlinear properties. The first approach has been extensively investigated by Madsen and co-workers with a first breakthrough in 1999 (Agnon \textit{et al.} 1999). In this work, the authors presented a new procedure by which it was possible to achieve the same accuracy on both linear and nonlinear properties. The main idea is to obtain approximate solutions to the Laplace equation (combined with the exact nonlinear free surface and bottom conditions) through truncated series expansions. While the formulation of Agnon \textit{et al.} (1999) involves velocity variables evaluated at the still-water level and is limited to $kh = 6$, the extended method proposed by Madsen \textit{et al.} (2002, 2003) completely removes the conventional shallow-water limitation, allowing for modelling fully nonlinear and highly dispersive waves up to $kh \approx 40$, i.e. in very deep water. This extended approach is based on velocity variables taken at optimized levels and optimal expansions through the use of Padé approximants. An extension to rapidly varying bathymetry has been proposed recently (Madsen \textit{et al.} 2006). Although a few numerical approaches based on this model have been proposed (Fuhrman \& Bingham 2004; Engsig-Karup \textit{et al.} 2006, 2008), the counterpart for this wide range of applicability is the numerical complexity of the underlying model which includes derivatives up to fifth-order and consists of more equations (and more unknowns) than the alternative model of Gobbi \textit{et al.} (2000). An interesting alternative approach has been chosen by Jamois \textit{et al.} (2006), where the authors propose to use a velocity potential formulation and to truncate the infinite series expansions of the solutions to the Laplace equation at a lower order: the resulting model is linearly and nonlinearly accurate only up to $kh = 10$, but entails a much lighter numerical complexity with less equations and with derivatives up to fourth-order only.

The second approach that has been studied is the double-layer approach, as proposed among others by Lynett \& Liu (2004$a$, $b$) and Audusse (2005). This approach is based on the idea of trading fewer unknowns and higher spatial derivatives for more unknowns and lower spatial derivatives. The multi-layering concept developed in the above references can be seen as an efficient mathematical tool to reduce the order of derivatives in any model, while increasing its linear and nonlinear range of applicability. The double-layer modelling proposed by Lynett \& Liu (2004$a$, $b$) is purely conceptual since the two layers have the same density. However, the resulting model, which is based on classical depth-integrated Boussinesq-type equations, allows to model accurately wave propagation up to $kh = 6$, both linearly and nonlinearly. This offers a very interesting alternative to high-order models, since this double-layer model is less complex (including derivatives up to third-order only) and more accurate in deep water.

The present work is mainly inspired by these two approaches, namely the
one of Madsen \textit{et al.} (2002, 2003) and the one of Lynett \& Liu
(2004$a$, $b$). The primary goal of this paper is to offer an efficient
alternative to the models of Madsen \textit{et al.} (2002, 2003) by
mixing their procedure, the simplifications of Jamois \textit{et al.}
(2006), and the double-layer concept of Lynett \& Liu (2004$a$, $b$). We
aim at deriving a model \textcolor{black}{which is 1) applicable to complex domains such
as coastal areas, islands or estuaries and 2)} linearly and nonlinearly accurate up to
deep water, but with lower complexity than the previous models
(i.e. lower order of derivatives and lower number of equations). The model
proposed herein satisfies all these conditions as it exhibits excellent
linear and nonlinear dispersive properties up to $kh=10$, consists of four equations 
in both one and two horizontal dimensions
(denoted by 1DH and 2DH, respectively), and includes up to second-order
spatial derivatives only. 

\textcolor{black}{
The present model hinges on a static Dirichlet-Neumann operator and
its approximation using a double-layer technique and Pad\'e
approximants. The advantage of using the static operator (that is,
defined on a fixed domain) as opposed to the usual Dirichlet-Neumann
operator defined at the free surface is that the static operator (or its
approximation) can be computed once and for all. The Dirichlet-Neumann
operator has been extensively investigated over the past two decades. 
Exact expressions of the static operator, thereby leading to exact
dispersion relations, can be found in the work of 
Craig \& Sulem (1993), Dommermuth
\& Yue (1987), Smith (1998), and in the work of
Matsuno (1993) and its extension to very general bathymetries by Artiles
\&  Nachbin (2004$a$, $b$). However, the application
of the above approaches to complex 2DH domains has not been reported
yet. Thus, with an eye toward coastal engineering applications, 
we prefer to base our approach on 
approximating the static Dirichlet-Neumann operator. 
Furthermore, we mention the new promising approach of Ablowitz \textit{et al.} 
(2006) based on an exact integral representation of the
usual Dirichlet-Neumann operator where no approximation is needed. 
Finally, we observe that the double-layer
technique used to derive the approximate static Dirichlet-Neumann
operator helps reducing the order of the derivatives while
improving the accuracy of the model. 
}

The paper is organized as follows. In \S 2, our model is formulated in terms of a static Dirichlet-Neumann operator, and we derive in \S 3 an approximation to this operator. A linear analysis of the model is presented in \S 4, including linear dispersion, vertical profiles of velocities, and linear shoaling. These properties are optimized based on Stokes linear wave theory, and it is shown that the model is accurate even for deep water conditions. Finally, in \S 5, numerical simulations are developed in 1DH to assess the nonlinear properties of the model for flat bottom conditions.

\section{Derivation of the double-layer formulation}

\subsection{Governing equations and boundary conditions}

\begin{figure}
\psfrag{z}{${\scriptstyle z}$}
\psfrag{horiz}{${\scriptstyle X=(x,y)}$}
\psfrag{zero}{${\scriptstyle 0}$}
\psfrag{omun}{${\scriptstyle \Omega_1}$}
\psfrag{omdeux}{${\scriptstyle \Omega_2}$}
\psfrag{eta}{${\scriptstyle z=\eta(t,X)}$}
\psfrag{inter}{${\scriptstyle z=\hat{z}(X)}$}
\psfrag{bot}{${\scriptstyle z=\bar{z}(X)}$}
\center{\includegraphics[width=10cm]{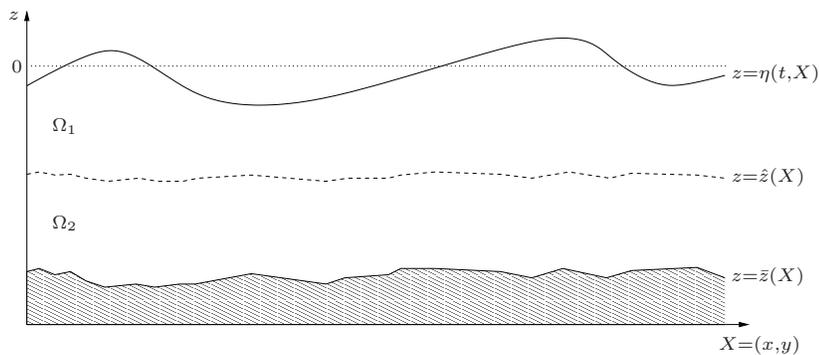}\hspace{2em}}
\caption{Representation of the fluid domain}
\label{figure1}
\end{figure}

We aim at formulating a double-layer Boussinesq-type model for the three-dimensional irrotational flow of an inviscid and incompressible fluid with a free surface. We focus here on so-called gravity waves or water waves, i.e. the evolution of a fluid under the only influence of gravity. The capillary effects due to the presence of surface tension are neglected. Moreover, we assume constant atmospheric pressure at the free surface of the fluid.
We adopt a Cartesian coordinate system, where we denote by $X=(x,y)$ the horizontal coordinates and by $z$ the vertical one, the $z$-axis pointing upwards. The time-dependent fluid domain is bounded from below by the static sea bottom and from above by the time-dependent free surface. We restrict this study to the case where the bathymetry and the free surface elevation are single-valued continuous functions, i.e. they can be described by the graph of two functions $X \mapsto \bar{z}(X) = -h(X)$ and $(t,X) \mapsto \eta(t,X)$ respectively. The level $z=0$ corresponds to the still-water level. As shown in figure \ref{figure1}, the fluid is divided into two layers by an interface $z = \hat{z}(X) = -\sigma h(X)$ where $\sigma$ is an arbitrary parameter in $]0,1[$. Thus, the thickness of the two layers are constant fractions of the still-water depth and do not depend on the free surface elevation. Unless the bottom is flat, the interface level $\hat{z}$ is therefore spatially (but not temporally) variable. The upper layer of the fluid is denoted by $\Omega_1$ and the lower layer by $\Omega_2$, namely $\Omega_1(t) = \{\,(X,z)\,;\;\hat{z}(X) \le z \le \eta(t,X)\}$ and $\Omega_2 = \{\,(X,z)\,; \bar{z}(X) \le z \le \hat{z}(X)\}$. We point out that this fluid division into two layers is purely conceptual since both layers have the same density.

As far as the bathymetry is concerned, we assume in this work that the still-water depth $h$ verifies $|\nabla h| \ll 1$, which corresponds to the classical mild-slope approximation. This approximation consists in neglecting all the quadratic (and higher order) terms in $\nabla h$ along with the derivatives of $h$ of order greater than one. Physically, this approximation means that we assume the wavelength of the free surface waves to be shorter than the distance over which the bathymetry (and thus the still-water depth) varies appreciably. \textcolor{black}{We point out that, in the mild-slope framework, the overall amplitude of
bottom topography levels can still be large. 
} The mild-slope approximation plays an important role in the derivation
and linear optimization of the present model, and it seems quite arduous
to incorporate higher-order bathymetric terms without significantly
increasing the complexity of the model. \textcolor{black}{For very general
  bathymetries in 1DH, we refer to Artiles \& Nachbin (2004$a$, $b$).} For clarity, all the equations derived using this mild-slope approximation are indicated in this work by the symbol $\ms$ instead of the equality symbol.

Since the flow is assumed irrotational, there exists a velocity potential $\phi$ such that ${\bf u} \equiv \nabla \phi$, $w \equiv \partial_z \phi$, where ${\bf u}$ denotes the horizontal velocity of the fluid, $w$ the vertical velocity, and $\nabla$ the horizontal gradient operator $(\partial_x,\,\partial_y)^T$. We define the velocity potentials $\phi_i$ and the vertical velocities $w_i$ in each layer by $\phi_i = \phi_{|_{\Omega_i}}\;,\;\;w_i = \partial_z \phi_i$ where the subscript $i \in \{1,2\}$ denotes the layer index.\\

The fluid motion in each layer is governed by the following equations written in terms of the velocity potential $\phi_i$ and the vertical velocity $w_i$,

\vspace{-1em}

\begin{subnumcases}{}
\Delta \phi_i + \partial_z^2 \phi_i = 0\,, \hspace{9.45em} \hspace{1em} (X,z) \in \Omega_i\;,\label{laplace}\\
\partial_t \phi_i + \frac{1}{2} |\nabla\phi_i|^2 + \frac{1}{2} w_i^2 + g z + P_i = R\,, \hspace{1em} (X,z) \in \Omega_i\;,\label{euler}
\end{subnumcases}
where $P_i$ is the reduced pressure field in each layer, $g$ the gravitational acceleration, and $R$ the Bernoulli constant. Equation (\ref{laplace}) corresponds to the Laplace (or continuity) equation and (\ref{euler}) corresponds to the Bernoulli (or momentum) equation. The Bernoulli constant $R$ only depends on time. Therefore, up to a time-dependent shift of the velocity potential, we can take this constant equal to $P_{atm}$ where $P_{atm}$ is the constant atmospheric pressure at the free surface.

\noindent At the free surface $z=\eta(t,X)$, the following boundary conditions are enforced:

\vspace{-1em}

\begin{subnumcases}{}
\partial_t \eta + \nabla\eta\cdot\nabla\phi_1 - w_1 = 0\,, \quad\text{at}\;z=\eta\;,\label{surface1}\\
P_1 = P_{atm}\,, \hspace{7.35em}\text{at}\;z=\eta\;,\label{surface2}
\end{subnumcases}

\noindent where (\ref{surface1}) is the usual kinematic free surface condition expressing that the free surface is a bounding surface, i.e. no fluid particle can cross it. At the interface $z=\hat{z}(X)$ between the layers, the following natural continuity conditions are enforced:
\begin{equation}\label{contcond}
\phi_1 = \phi_2\,,\quad w_1 = w_2\,,\quad P_1 = P_2\,,\quad\,\text{at}\;z=\hat{z}\;.
\end{equation}
Observe that $\phi_1 = \phi_2$ and $w_1 = w_2$ at $z=\hat{z}$ imply $\nabla\phi_1 = \nabla\phi_2$ and thus $u_1 = u_2$ at $z=\hat{z}$, hence recovering the continuity conditions on the horizontal and vertical velocities enforced by Lynett \& Liu (2004$a$).

\noindent Finally, the system of equations is closed by the usual kinematic condition at the sea bottom $z=\bar{z}(X)$:
\begin{equation}\label{fond}
\nabla h \cdot\nabla\phi_2 + w_2 = 0\,,\quad\text{at}\;z=\bar{z}\;,
\end{equation}
which expresses that the sea bottom is a bounding surface. We now introduce
$$
\left\{
\begin{array}{lcl}
\vspace{0.2em}
\widetilde{\phi_1}(t,X) = \phi_1(t,X,z=\eta(t,X))\,, & & \;\;\widetilde{w_1} = w_1(t,X,z=\eta(t,X))\;,\\
\vspace{0.2em}
\widehat{\phi_i}(t,X) = \phi_i(t,X,z=\hat{z}(X))\,, & & \;\;\widehat{w_i} = w_i(t,X,z=\hat{z}(X))\;,\\
\overline{\phi_2}(t,X) = \phi_2(t,X,z=\bar{z}(X))\,, & & \;\;\overline{w_2} = w_2(t,X,z=\bar{z}(X))\;.
\end{array}
\right.
$$
Following Zakharov (1968), we can reformulate the equations as

\vspace{-1.25em}

\begin{subequations}
\begin{gather}
\partial_t \widetilde{\phi_1} + \frac{1}{2} |\nabla\widetilde{\phi_1}|^2 - \frac{1}{2}\widetilde{w_1}^2 (1+|\nabla\eta|^2) + g\eta = 0\;, \label{fin1}\\
\partial_t \eta + \nabla\eta\cdot\nabla\widetilde{\phi_1}-\widetilde{w_1} (1+|\nabla\eta|^2) = 0\;, \label{fin2}
\end{gather}
\noindent where (\ref{fin1}) is the Euler equation expressed at the free surface and (\ref{fin2}) the kinematic condition at the free surface,

\vspace{-1.5em}

\begin{gather}
\Delta \phi_1 + \partial_z w_1 = 0\;, \quad \hat{z} \le z \le \eta\;,\color{white}{\frac{1}{2}} \label{fin3a}\\
\Delta \phi_2 + \partial_z w_2 = 0\;, \quad \bar{z} \le z \le \hat{z}\;,\color{white}{1}\, \label{fin4a}
\end{gather}

\noindent where (\ref{fin3a}) and (\ref{fin4a}) are the Laplace equations in each layer, and

\vspace{-1.5em}

\begin{gather}
\widehat{\phi_1} = \widehat{\phi_2}\;,\color{white}{\frac{1}{2}}\;\, \label{fin5}\\
\widehat{w_1} = \widehat{w_2}\;,\color{white}{1_{|_2}} \label{fin6}\\
\overline{w_2} + \nabla h \cdot \nabla\overline{\phi_2} \ms 0\;,\color{white}{1}\, \label{fin7}
\end{gather}
\end{subequations}
where (\ref{fin5}) and (\ref{fin6}) correspond to the continuity conditions at the interface, and (\ref{fin7}) to the kinematic condition at the bottom, where we used the mild-slope hypothesis to neglect the $|\nabla h|^2$ term. We point out that we work with a velocity potential formulation, unlike Madsen \textit{et al.} (2002, 2003) who formulated the governing equations in terms of the velocity variables ${\bf u}$ and $w$. This choice stems from our will to minimize the total number of equations in the model: in 2DH and in the present double-layer framework under the irrotational flow assumption, our velocity potential formulation allows to consider two less equations than with a velocity formulation.

In the sequel, equations (\ref{fin1}) and (\ref{fin2}) are left unchanged as they define the fully nonlinear time-stepping problem. We focus on the Laplace equations and the remaining boundary conditions to close the time-stepping problem by expressing the vertical velocity at the free surface $\widetilde{w_1}$ in terms of the velocity potential $\widetilde{\phi_1}$ at that surface, the free surface $\eta$, and the bathymetry $h$. This relation corresponds to the well-known Dirichlet-Neumann operator. The next subsections and \S 3 are devoted to the crucial construction of an accurate, yet computationally cheap approximation to this operator.

\subsection{A translated Dirichlet-Neumann operator}

The Dirichlet-Neumann operator associated to the problem (\ref{fin3a})--(\ref{fin7}) is denoted by $\mathcal{G}[\eta,h]$ and is defined by $\mathcal{G}[\eta,h] \psi = \partial_z \phi_1|_{z=\eta}$ for any smooth enough function $\psi$, where $(\phi_1,\phi_2)$ solves the boundary value problem composed of equations (\ref{fin3a})--(\ref{fin7}) along with the Dirichlet condition $\phi_1 = \psi$ at $z = \eta$. One can thus simply write the closure between the unknowns $\widetilde{w_1}$, $\widetilde{\phi_1}$, and $\eta$ as
\begin{equation}\label{DN3}
\widetilde{w_1} = \mathcal{G}[\eta,h] \widetilde{\phi_1}\;.
\end{equation}
This Dirichlet-Neumann operator is at the heart of the derivation of Boussinesq-type models since the structure and accuracy of these models essentially depend on the method used to construct an approximation to this operator. Once this approximation is derived, there are two options. The first one is to eliminate the vertical velocity variable $\widetilde{w_1}$ from the equations by plugging (\ref{DN3}) into (\ref{fin1}), (\ref{fin2}). This method has been classically used in Boussinesq-type models and has the advantage of lowering the number of equations to solve at each time step, but significantly increases their complexity. The other option has been used for instance by Madsen \textit{et al.} (2002, 2003) and consists in keeping $\widetilde{w_1}$ in the equations, which entails to solve (\ref{fin1}), (\ref{fin2}) and then to compute $\widetilde{w_1}$ through the use of the Dirichlet-Neumann operator $\mathcal{G}[\eta,h]$ at each time step. This is the method we have chosen to use to keep equations complexity to a minimum.

The main difficulty in finding an approximation to the Dirichlet-Neumann operator is that it involves solving the Laplace equations (\ref{fin3a}) and (\ref{fin4a}) along with the boundary conditions (\ref{fin5})--(\ref{fin7}) and $\phi_1 = \widetilde{\phi_1}$ at $z = \eta$, on a time-dependent domain bounded from above by the free surface $z=\eta$. Keeping $\widetilde{w_1}$ in the equations involves constructing an approximation to $\mathcal{G}[\eta,h]$ at each time step, which can be a serious computational problem as it increases the numerical cost. An interesting work-around to this issue consists in constructing an alternative Dirichlet-Neumann operator expressed at the still-water level, and then finding a closure between the unknown functions at the free surface $z=\eta$ and the ones at the still-water level $z=0$. To this end, we introduce $\phi_0(t,X) = \phi_1(t,X,z=0)$ and $w_0(t,X) = w_1(t,X,z=0)$
and denote by $\mathcal{G}_0[h]$ the Dirichlet-Neumann operator corresponding to $\eta=0$, i.e.
\begin{equation}\label{TDNdef}
\mathcal{G}_0[h] = \mathcal{G}[0,h]\;.
\end{equation}

\noindent The translated operator $\mathcal{G}_0[h]$ is such that $\mathcal{G}_0[h] \psi = \partial_z \phi_1|_{z=0} = w_0$ where $(\phi_1,\phi_2)$ solves the boundary value problem
\begin{equation}\label{DN4}
\left\{
\begin{array}{ll}
\Delta \phi_1 + \partial_z^2 \phi_1 = 0\,, & \quad
\hat{z} \le z \le 0\;,\\
\Delta \phi_2 + \partial_z^2 \phi_2 = 0\,, & \quad
\bar{z} \le z \le \hat{z}\;,\\
\phi_1 = \psi\,, & \quad \text{at}\; z = 0\;,\\
\phi_1 = \phi_2\,, \quad \partial_z \phi_1 = \partial_z \phi_2\,, & \quad \text{at}\; z=\hat{z}\;,\\
\partial_z \phi_2 + \nabla h \cdot \nabla\phi_2 = 0\,, & \quad \text{at}\; z=\bar{z}\;.
\end{array}
\right.
\end{equation}

\noindent The main advantage of this translated Dirichlet-Neumann operator is that the boundary value problem (\ref{DN4}) is posed on a static domain bounded from above by the still-water level $z=0$ and from below by the sea bottom $z=\bar{z}$. 

\begin{rmrk}
We point out that $\phi_0$ is not defined everywhere since the bounding free surface $\eta$ can take negative values. This issue can be solved by extending the solutions to the Laplace equation above $z=\eta$ when $\eta < 0$, using the fact that both $\phi_1$ and its normal derivative $w_1$ are continuous at the free surface. This mathematical trick allows to artificially define $\phi_0$ and $w_0$ when $\eta < 0$.  This tool has been implicitly used by many authors such as Nwogu (1993), Wei \textit{et al.} (1995), Gobbi \textit{et al.} (2000) or Madsen \textit{et al.} (2002, 2003) who derived models based on an horizontal velocity (or potential) variable taken at free-surface independent levels, thus allowing these levels to exceed the bounding value $z=\eta$ for $\eta$ negative enough.
\end{rmrk}

\subsection{Closure relation and model formulation}

Our goal is twofold. Firstly to construct an approximation to the translated Dirichlet-Neumann operator $\mathcal{G}_0[h]$, and secondly to look for closure relations between the unknowns $\widetilde{\phi_1}$, $\widetilde{w_1}$, $\phi_0$, and $w_0$. The second objective can be readily achieved via a Taylor expansion of $\phi_1$ and $w_1$ at the still-water level $z=0$. Indeed, combining the MacLaurin expansions of $\phi_1$ and $w_1$ at respectively the fourth and third order (see Remark \ref{truncrmrk}) and the Laplace equation $\Delta \phi_1 = -\partial_z w_1$ at $z=0$ yields the desired closure relations, namely
\begin{equation}\label{closure}
\left\{
\begin{array}{rcl}
\vspace{0.25em}
\hspace{-0.25em}\widetilde{\phi_1} & \hspace{-0.5em}= & \hspace{-0.5em}(1-\widetilde{\alpha_1}\Delta)\, \phi_0 + (\widetilde{\beta_1}-\widetilde{\gamma_1}\Delta)\, w_0\;,\\
\hspace{-0.25em}\widetilde{w_1} & \hspace{-0.5em}= & \hspace{-0.5em}-\widetilde{\beta_1} \Delta\, \phi_0 + (1-\widetilde{\alpha_1}\Delta)\, w_0\;,
\end{array}
\right.
\end{equation}
where $\displaystyle{\widetilde{\alpha_1} = \frac{\eta^2}{2}}$, $\displaystyle{\widetilde{\beta_1} = \eta}$, and $\displaystyle{\widetilde{\gamma_1} = \frac{\eta^3}{6}}$.

\vspace{0.5em}

\noindent We can now state our double-layer model as follows:
\begin{equation} \label{model}
\left\{
\begin{array}{l}
\vspace{0.25em}
\partial_t \widetilde{\phi_1} + \frac{1}{2} |\nabla\widetilde{\phi_1}|^2 - \frac{1}{2}\widetilde{w_1}^2 (1+|\nabla\eta|^2) + g\eta = 0\;,\\
\vspace{0.25em}
\partial_t \eta + \nabla\eta\cdot\nabla\widetilde{\phi_1}-\widetilde{w_1} (1+|\nabla\eta|^2) = 0\;,\\
\vspace{0.25em}
\left( 1-\frac{\eta^2}{2}\Delta + (\eta - \frac{\eta^3}{6}\Delta)\,\mathcal{G}_0^{app\,}[h] \right) \phi_0 = \widetilde{\phi_1}\;,\\
\widetilde{w_1} = \left(- \eta \Delta + (1-\frac{\eta^2}{2}\Delta)\,\mathcal{G}_0^{app\,}[h] \right) \phi_0\;.
\end{array}
\right.
\end{equation}
where $\mathcal{G}_0^{app\,}[h]$ is an approximation to the static Dirichlet-Neumann operator $\mathcal{G}_0[h]$ which is detailed in the following section.\\

The main advantages of this model are that 1) it consists of only four
equations both in 1DH and 2DH, \textcolor{black}{2) it can be used in
  complex domains}, and 3) the approximate Dirichlet-Neumann operator $\mathcal{G}_0^{app\,}[h]$ can be computed once and for all at $t=0$ since this operator is static. Furthermore, we will see in \S 3 that it only includes at most second-order horizontal derivatives. This is a major improvement in comparison with high-order Boussinesq-type models such as those of Jamois \textit{et al.} (2006) and Madsen \textit{et al.} (2002, 2003) which contain respectively fourth- and fifth-order derivatives, and consist of respectively five equations in 1DH and 2DH, and five equations in 1DH and seven in 2DH.

\vspace{0.25em}

\begin{rmrk}\label{truncrmrk}
The truncation orders used in (\ref{closure}) can be motivated by a dimensional analysis. We scale the vertical coordinate $z$ and the free surface $\eta$ with the typical amplitude $a$ of the surface waves, the horizontal coordinate $X$ by the typical wavelength $\lambda$ and introduce the mean depth $h_0$. The truncation errors in the two equations of (\ref{closure}) are respectively of order $O(\varepsilon^4\mu^2,\varepsilon^5\mu^2)$ and $O(\varepsilon^4\mu^2,\varepsilon^3\mu^2)$, where $\varepsilon = a/h_0$ and $\mu = {h_0^2}/{\lambda^2}$ correspond respectively to the nonlinearity and dispersion parameters. Thus, the truncated terms are almost third and fourth powers of the steepness parameter $s$ defined by $s=\varepsilon\sqrt{\mu}$, whose maximum value is admittedly $s_{max} \approx 0.142$ for a stable wave (Williams 1981). Combining this value and the fact that we consider fully nonlinear waves, for which $\varepsilon$ is of order $O(1)$, motivates the truncation order in (\ref{closure}).
\end{rmrk}

\section{An approximate static Dirichlet-Neumann operator}

\subsection{Theoretical solutions to the Laplace equations}

The first step in the derivation of the approximate static Dirichlet-Neumann operator $\mathcal{G}_0^{app}[h]$ is to look for solutions to the Laplace equations
\begin{equation}\label{fin34}
\Delta\phi_i + \partial^2_z \phi_i = 0\;,\quad (X,z) \in \Omega_i\;,
\end{equation}
where we have redefined the upper-layer domain as $\Omega_1 = \{\,(X,z)\,;\;\hat{z}(X) \le z \le 0\}$. To this end, we follow the generalized Boussinesq procedure introduced by Madsen \textit{et al.} (2002, 2003) which consists in looking for a solution under the form of an infinite Taylor series in the vertical coordinate. The main difference between this method and the classical Boussinesq procedure is that in the latter, one looks for a finite series solution in the vertical coordinate, i.e. a low-order polynomial in the variable $z$. The generalized method of Madsen \textit{et al.} (2002, 2003) allows to find exact infinite series solutions instead of approximate solutions.\\
We first introduce two arbitrary expansion levels $z_1$ and $z_2$ in each layer, namely $z_1(X) = - \sigma_1\,h(X)$ with $0 < \sigma_1 < \sigma$ and 
$z_2(X) = - \sigma_2\,h(X)$ with $\sigma < \sigma_2 < 1$, and the associated unknowns $\breve{\phi_i}$ and $\breve{w_i}$ such that $\breve{\phi_i} = \phi_i(X,z=z_i(X))$ and $\breve{w_i} = w_i(X,z=z_i(X))$ for $i \in \{1,2\}$. We now look for solutions to the Laplace equations in the form of infinite Taylor series in the vertical variable $(z-z_i)$,
\begin{equation}\label{ansatz}
\phi_i(X,z) = \sum_{n \ge 0}{(z-z_i)^n \phi_{\;i}^{(n)}(X)}\;,
\end{equation}
where the choice of the vertical variable $(z-z_i)$ instead of $z$ actually allows to save one step compared to the procedure of Madsen \textit{et al.} (2002, 2003). Injecting (\ref{ansatz}) into the Laplace equations (\ref{fin34}) and using the mild-slope assumption leads to the recurrence relation
\begin{equation}\label{zz}
\Delta \phi_{\;i}^{(n)} - 2 (n+1) \nabla z_i \cdot \nabla\phi_{\;i}^{(n+1)} + (n+2)(n+1) \phi_{\;i}^{(n+2)} \ms 0\;.
\end{equation}
Observing that $\Delta(\nabla z_i \cdot \nabla \phi_{\;i}^{(k)}) \ms \nabla z_i \cdot \Delta \nabla \phi_{\;i}^{(k)}$ for all $k$ yields the expression of $\phi_{\;i}^{(2n)}$ and $\phi_{\;i}^{(2n+1)}$ in terms of $\breve{\phi_i}$ and $\breve{w_i}$,
$$
\left\{
\begin{array}{l}
\phi_{\;i}^{(2n)} \ms \frac{(-1)^n}{(2n)!}\Delta^n \breve{\phi_i} + \frac{(-1)^{n-1}}{(2n-1)!}\nabla z_i \cdot \Delta^{n-1}\nabla \breve{w_i}\;,\\
\phi_{\;i}^{(2n+1)} \ms \frac{(-1)^n}{(2n+1)!}\Delta^n \breve{w_i} + (-1)^{n}\frac{2n}{(2n+1)!}\nabla z_i \cdot \Delta^{n}\nabla \breve{\phi_i}\;.
\end{array}
\right.
$$
Plugging these expressions into the ansatz (\ref{ansatz}) provides the desired expressions for the velocity potentials $\phi_i$ and vertical velocity $w_i$ in terms of $\breve{\phi_i}$ and $\breve{w_i}$ for all $(X,z)$ belonging to $\Omega_i$,
\begin{equation}\label{complete}
\left\{
\begin{array}{l}
\vspace{0.25em}
\phi_i(X,z) = \cot(z-z_i) \breve{\phi_i} + \sit(z-z_i) \breve{w_i} + \nabla z_i \cdot \Gamma_{\phi_i}\;,\\
w_i(X,z) = - \sit(z-z_i) \Delta \breve{\phi_i} + \cot(z-z_i) \breve{w_i} + \nabla z_i \cdot \Gamma_{w_i}\;,
\end{array}
\right.
\end{equation}
where $\cot$ and $\sit$ are infinite-series pseudo-differential operators defined by
\begin{equation}\label{cossin}
\cot(\lambda) = \sum_{n \ge 0} (-1)^n \frac{\lambda^{2n}}{(2n)!} \Delta^{n}\;,\quad
\sit(\lambda) = \sum_{n \ge 0} (-1)^n \frac{\lambda^{2n+1}}{(2n+1)!} \Delta^{n}\;,
\end{equation}
and where the slope terms $\Gamma_{\phi_i}$ and $\Gamma_{w_i}$ are given by
\begin{equation}\label{gamma}
\left\{
\begin{array}{l}
\vspace{0.25em}
\Gamma_{\phi_i} = (z-z_i)\left[\,\cot(z-z_i) \nabla \breve{\phi_i} + \sit(z-z_i) \nabla \breve{w_i}\right] - \sit(z-z_i) \nabla \breve{\phi_i}\;,\\
\Gamma_{w_i} = (z-z_i)\left[-\sit(z-z_i)\nabla\Delta \breve{\phi_i} + \cot(z-z_i) \nabla \breve{w_i}\right] + \sit(z-z_i) \nabla \breve{w_i}\;.
\end{array}
\right.
\end{equation}
The expression (\ref{complete}) of $\phi_i$ provides a theoretical formulation of an exact solution to the Laplace equations (\ref{fin34}). Strictly speaking, we can verify that they are in fact solutions to (\ref{fin34}) with residuals of order $O(|\nabla h|^2, \Delta h)$ which are negligible within our mild-slope approximation framework.

\subsection{Truncation of the Taylor series}

The previous solutions to the Laplace equations (\ref{fin34}) are purely theoretical since they involve infinite-series pseudo-differential operators. To deal with this problem, we obviously need to truncate the series at a finite order, and this raises the question of choosing the order of truncation. Through this choice, we have to reach a compromise between the accuracy of the truncated expression and the numerical complexity of the final model. In fact, the truncation order is a key factor for the domain of validity of the model: the higher the truncation order is, the better dispersive effects are reproduced, so that the model is applicable in deeper water. We recover here the common paradigm encountered in works based on asymptotic expansions of the solutions to (\ref{fin34}) where retaining higher order terms increases the domain of validity in intermediate or deep water.

Within our double-layer framework, the increase of the number of unknowns allows us to lower the truncation order in comparison with the works of Madsen \textit{et al.} (2002, 2003), and even of Jamois \textit{et al.} (2006). We take advantage of this possibility, and truncate the operators $\cot$ and $\sit$ by retaining only the first two terms of the series, which leads to the following approximations
\begin{equation}\label{cossint}
\cot(\lambda) = 1-\frac{\lambda^2}{2}\Delta + O(\lambda^4\Delta^2)\;,\quad
\sit(\lambda) = \lambda-\frac{\lambda^3}{6}\Delta + O(\lambda^5\Delta^2)\;.
\end{equation}
We first plug these approximations into (\ref{complete}), which leads to the following truncated expressions of $\phi_i$ and $w_i$
\begin{equation}\label{trunc}
\left\{
\begin{array}{rcl}
\vspace{0.25em}
\hspace{-0.25em}\phi_i(X,z) & \hspace{-0.5em}= & \hspace{-0.5em}(1-\alpha_i\Delta) \breve{\phi_i} + (\beta_i-\gamma_i\Delta) \breve{w_i}
+ \nabla z_i \cdot \left[ \beta_i \left( (\nabla - \alpha_i\nabla\Delta) \breve{\phi_i} \right.\right.\\
& & \hspace{-0.25em}\left. + (\beta_i\nabla-\gamma_i\nabla\Delta) \breve{w_i} \Big) 
- (\beta_i\nabla - \gamma_i \nabla \Delta) \breve{\phi_i} \right]\,,\\
\vspace{0.25em}
\hspace{-0.25em}w_i(X,z) & \hspace{-0.5em}= & \hspace{-0.5em}(-\beta_i \Delta + \gamma_i \Delta^2) \breve{\phi_i} + (1-\alpha_i\Delta) \breve{w_i}
+ \nabla z_i \cdot  \Big[ \beta_i \Big( (-\beta_i\nabla\Delta \\
& & \hspace{-0.25em}+ \gamma_i\nabla\Delta^2) \breve{\phi_i} + (\nabla - \alpha_i\nabla\Delta) \breve{w_i} \Big) + (\beta_i\nabla - \gamma_i\nabla\Delta) \breve{w_i} \Big]\,,
\end{array}
\right.
\end{equation}
where
\begin{equation}\label{coeffs}
\alpha_i(z) = \frac{(z-z_i)^2}{2}\;\;,\;\;\beta_i(z) = z-z_i\;\;,\;\;\gamma_i(z) = \frac{(z-z_i)^3}{6}\;.
\end{equation}
We now reformulate these expressions by applying the operators $P_i$, defined for all smooth enough scalar-valued function $u$ by $P_i\,u = u - \beta_i \nabla z_i \cdot \nabla u$, which yields the following expressions for all $(X,z)$ in $\Omega_i$,
\begin{equation}\label{trunc2}
\left\{
\begin{array}{l}
\vspace{0.5em}
\hspace{-0.45em}P_i \phi_i \hspace{0.25em}\ms\hspace{0.25em} (1-\alpha_i\Delta) \breve{\phi_i} + (\beta_i-\gamma_i\Delta) \breve{w_i} - \nabla z_i \cdot \left[ (\beta_i\nabla - \gamma_i \nabla \Delta) \breve{\phi_i} \right]\,,\\
\hspace{-0.45em}P_i w_i \hspace{0.25em}\ms\hspace{0.25em} (-\beta_i \Delta + \gamma_i \Delta^2) \breve{\phi_i} + (1-\alpha_i\Delta) \breve{w_i} + \nabla z_i \cdot \Big[ (\beta_i\nabla - \gamma_i\nabla\Delta) \breve{w_i} \Big]\,.
\end{array}
\right.
\end{equation}
The advantage of this new formulation will become clear in the sequel.

\begin{rmrk}
The chosen order of truncation can be motivated as in remark \ref{truncrmrk} by a dimensional analysis in the case of a flat bottom. Scaling here $z$ and the expansion levels $z_i$ with $h_1 = \sigma h_0$ in the upper-layer and $h_2 = (1-\sigma) h_0$ in the lower-layer, the dimensionless form of (\ref{complete})--(\ref{cossin}) is obtained by replacing $\lambda^2$ by $\mu_i$ in (\ref{cossin}), where $\mu_i = {h_i^2}/{\lambda^2}$ corresponds to the dispersion parameter in each layer. Supposing that all the derivatives are of order $O(1)$, we can analyze the order of the third terms of each series in deep water, for instance $k h_0 = 10$, where $\mu_i$ is considerably higher than in shallow water. If we restrict
$\sigma$ to the range $[0.25,0.75]$, this leads to the following estimates for $n=2$ in deep water:
$\mu_i^{n}/((2n)!) \approx 0.08$, $\mu_i^{n}/((2n+1)!) \approx 0.01$ and $\mu_i^{n+1}/((2n+1)!) \approx 0.02$.
Thus, the third term of each infinite series in (\ref{complete})--(\ref{cossin}) is very small in deep water: these terms and all the subsequent ones can be neglected. The same kind of analysis performed for an uneven bottom leads to the same result. This motivates the truncation order chosen in (\ref{trunc}).
\end{rmrk}

\subsection{Padé approximants}

Using the truncated expression (\ref{trunc2}), it is possible to construct an approximation to the static Dirichlet-Neumann operator, but involving up to fourth-order differential operators at first order in $h$ and fifth-order differential operators in the slope terms. Consequently, we now present a method to lower the maximum order of the derivatives in (\ref{trunc2}) based on Padé approximants. We follow the strategy introduced by Madsen \textit{et al.} (2002, 2003) and expand the variables $\breve{\phi_i}$ and $\breve{w_i}$ in terms of auxiliary variables $\phi_i^{\ast}$ and $w_i^{\ast}$ through the relations
\begin{equation}
\left\{
\begin{array}{l}
\vspace{0.25em}
\breve{\phi_i} = M_i(z_i\nabla)\,\phi_i^{\ast}\,,\;\;
\breve{w_i} = M_i(z_i\nabla)\,w_i^{\ast}\,,\\
M_i(z_i\nabla) = 1 + p_i z_i^2 \Delta + q_i z_i \nabla z_i \cdot \nabla\,,
\end{array}
\right.
\end{equation}
where $p_i$ and $q_i$ are arbitrary coefficients to be determined.\\
We now plug this ansatz into (\ref{trunc2}) and conduct the same dimensional analysis as previously, which yields the following expressions:
\begin{equation}\label{padé}
\hspace{-0.25em}\left\{
\begin{array}{rcl}
\vspace{0.15em}
\hspace{-0.5em}P_i \phi_i & \hspace{-0.5em}\ms & \hspace{-0.5em}\left(1 - (\alpha_i - p_i z_i^2) \Delta \right) \phi_i^{\ast} + \left(\beta_i - (\gamma_i - \beta_i p_i z_i^2) \Delta \right) w_i^{\ast}\\
\vspace{0.15em}
& & \hspace{-0.5em}+ \nabla z_i \cdot \Big[ \left((q_i z_i - \beta_i) \nabla + (\gamma_i - \beta_i p_i z_i^2 - \alpha_i(q_i+4p_i)z_i) \nabla\Delta \right) \phi_i^{\ast}\\
\vspace{0.15em}
& & \hspace{-0.5em}+ \left(\beta_i q_i z_i \nabla - \gamma_i(q_i+4p_i)z_i \nabla\Delta \right) w_i^{\ast} \Big]\;,\\
\vspace{0.15em}
\hspace{-0.5em}P_i w_i & \hspace{-0.5em}\ms & \hspace{-0.5em}\left(-\beta_i \Delta + (\gamma_i - \beta_i p_i z_i^2) \Delta^2 \right) \phi_i^{\ast} + \left(1 - (\alpha_i - p_i z_i^2) \Delta \right) w_i^{\ast}\\
\vspace{0.15em}
& & \hspace{-0.5em}+ \nabla z_i \cdot \Big[ \left(-\beta_i(q_i+4p_i)z_i \nabla\Delta \right) \phi_i^{\ast} + \left((\beta_i + q_i z_i) \nabla \textcolor{white}{z_i^2} \right.\\ 
& & \hspace{-0.5em}\left. - (\gamma_i - \beta_i p_i z_i^2 + \alpha_i(q_i+4p_i)z_i) \nabla\Delta \right) w_i^{\ast} \Big]\;,
\end{array}
\right.
\end{equation}
where we have again kept the first two terms in each modified truncated series, except the fifth-derivative of $\phi_i^{\ast}$ appearing in the slope terms of $P_i w_i$. This choice is motivated by the mild-slope approximation, but it clearly unbalances the global structure of the slope terms and does have a negative impact on the linear shoaling. However, we present in \S 4 a remedy to this problem.\\

We now aim at lowering the maximum order of the derivatives in (\ref{padé}) while preserving the overall accuracy of the truncated expressions (\ref{trunc2}). This goal can be achieved by choosing the coefficients $p_i$ and $q_i$  in order to introduce Padé approximants in the equations. In Madsen \textit{et al.} (2002, 2003), the authors use Padé approximants as a way to improve truncation accuracy without increasing the order of the derivatives. In the present work, we rather view the Padé approximants as a means to cancel high-order derivatives in (\ref{padé}) while preserving the accuracy of the truncated expressions (\ref{trunc2}).

Practically, lowering the maximum derivative order in (\ref{padé}) means requiring that the factors $\gamma_i - \beta_i p_i z_i^2$ and $q_i+4p_i$ (respectively in front of the fourth-order and third-order derivatives) vanish in each layer, via an appropriate choice of the constants $p_i$, $q_i$, and $\sigma_i$ (that define the expansion levels $z_i$). Since the quantity $\gamma_i - \beta_i p_i z_i^2$ depends on the vertical variable $z$ through $\gamma_i$ and $\beta_i$, it is impossible that this factor vanishes over the whole still-water column. Nevertheless, the truncated expressions (\ref{trunc2}) of $P_i \phi_i$ and $P_i w_i$ need to be evaluated only at some levels, namely $z=0$ for the upper boundary of $\Omega_1$, $z=\hat{z}$ for the interface and $z=\bar{z}$ for the sea bottom. Consequently, requiring that the quantity $\gamma_1 - \beta_1 p_1 z_1^2$ vanishes at the still-water level and at the interface, and that the quantity $\gamma_2 - \beta_2 p_2 z_2^2$ vanishes at the interface and at the sea bottom eliminates all the fourth-order derivatives from the model. Using (\ref{coeffs}), this yields
$$
\left\{
\begin{array}{l}
\vspace{0.5em}
p_1 = \frac{1}{6}\,,\;\;p_2 = \frac{1}{6}\left(\frac{1-\sigma}{1+\sigma}\right)^2\,,\\
\sigma_1 = \frac{\sigma}{2}\,,\;\;\sigma_2 = \frac{\sigma+1}{2}\,.
\end{array}
\right.
$$
Then, taking $q_i = -4 p_i$ yields $q_1 = -\frac{2}{3}$ and $q_2 = -\frac{2}{3}\left(\frac{1-\sigma}{1+\sigma}\right)^2$.
We observe that the expansion levels $z_1$ and $z_2$, respectively defined by $z_1 = - \sigma_1 h$ and $z_2 = - \sigma_2 h$, are thus taken respectively at the middle of the upper layer at rest and at the middle of the lower one.\\

We can now plug the previous values for $p_i$, $q_i$, and $z_i$ into the expressions (\ref{padé}) of $P_i\phi_i$ and $P_i w_i$, and evaluate them at the three boundary levels $z=0$, $z=\hat{z}$, and $z=\bar{z}$. We first obtain the following expressions at the still-water level:
\begin{equation}\label{surf}
\hspace{-0.1em}\left\{
\begin{array}{ll}
\vspace{0.35em}
\hspace{-0.5em}\left( 1 + \frac{\sigma}{2}{\beta_1^{\ast}} \nabla h \cdot \!\nabla \right) {\phi_0} = \left( 1 - {\alpha_1^{\ast}} \Delta \right) \phi_1^{\ast} + {\beta_1^{\ast}} w_1^{\ast} + \nabla h \cdot \! \Big[ {\gamma_1^{\ast}} \nabla \phi_1^{\ast} - {\delta_1^{\ast}} \nabla w_1^{\ast} \Big],\\
\hspace{-0.5em}\left( 1 + \frac{\sigma}{2}{\beta_1^{\ast}} \nabla h \cdot \!\nabla \right) {w_0} = - {\beta_1^{\ast}} \Delta \phi_1^{\ast} + \left( 1 - {\alpha_1^{\ast}} \Delta \right) w_1^{\ast}
- {\varepsilon_1^{\ast}} \nabla h \cdot \!\nabla w_1^{\ast},
\end{array}
\right.
\end{equation}
where
\begin{equation}\label{coeff1}
\alpha_1^{\ast} = \frac{\sigma^2}{12} h^2\,,\quad
\beta_1^{\ast} = \frac{\sigma}{2} h\,,\quad
\gamma_1^{\ast} = \frac{\sigma^2}{12} h\,,\quad
\delta_1^{\ast} = \frac{\sigma^3}{12} h^2\,,\quad
\varepsilon_1^{\ast} = \frac{5\sigma^2}{12} h\,.
\end{equation}
At the interface $z=\hat{z}$, we use the mild-slope approximation to obtain
\begin{equation}\label{interf}
\hspace{-0pt}\left\{
\!\!\!
\begin{array}{l}
\vspace{0.45em}
\hspace{-0.25em} \left( 1 - {\textstyle\frac{\sigma}{2}}\beta_1^{\ast} \nabla h \cdot\! \nabla \right) \widehat{\phi_1} \ms \left( 1 - \alpha_1^{\ast} \Delta \right) \phi_1^{\ast} - \beta_1^{\ast} w_1^{\ast} + \nabla h \cdot\! [ - \varepsilon_1^{\ast} \nabla \phi_1^{\ast} + \delta_1^{\ast} \nabla w_1^{\ast} ],\\
\vspace{0.45em}
\hspace{-0.25em} \left( 1 - {\textstyle\frac{\sigma}{2}}\beta_1^{\ast} \nabla h \cdot\! \nabla \right) \widehat{w_1} \ms \beta_1^{\ast} \Delta \phi_1^{\ast} + \left( 1 - \alpha_1^{\ast} \Delta \right) w_1^{\ast}
+ \nabla h \cdot\! [ \gamma_1^{\ast} \nabla w_1^{\ast} ],\\
\vspace{0.45em}
\hspace{-0.25em} \Big( 1 + {\textstyle\frac{\sigma+1}{2}}\beta_2^{\ast} \nabla h \!\cdot\! \nabla \Big) \, \widehat{\phi_2} \ms \left( 1 - \alpha_2^{\ast} \Delta \right) \phi_2^{\ast} + \beta_2^{\ast} w_2^{\ast} + \nabla h \!\cdot\! [ \gamma_2^{\ast} \nabla \phi_2^{\ast} - \delta_2^{\ast} \nabla w_2^{\ast} ],\\
\hspace{-0.25em} \Big( 1 + {\textstyle\frac{\sigma+1}{2}}\beta_2^{\ast} \nabla h \cdot\! \nabla \Big) \, \widehat{w_2} \ms - \beta_2^{\ast} \Delta \phi_2^{\ast} + \left( 1 - \alpha_2^{\ast} \Delta \right) w_2^{\ast}
- \nabla h \cdot\! [ \varepsilon_2^{\ast} \nabla w_2^{\ast} ],
\end{array}
\right.
\end{equation}
where
\begin{equation}\label{coeff2}
\begin{array}{c}
\alpha_2^{\ast} = \frac{(1-\sigma)^2}{12} h^2\;,\quad
\beta_2^{\ast} = \frac{1-\sigma}{2} h\;,\quad
\gamma_2^{\ast} = \frac{5\sigma+1}{12} (1-\sigma) h\;,\\
\delta_2^{\ast} = \frac{(1-\sigma)^3}{12} h^2\;,\quad
\varepsilon_2^{\ast} = \frac{\sigma+5}{12} (1-\sigma) h\;.
\end{array}
\end{equation}
Finally, the expressions at the sea bottom $z=\bar{z}$ are
\begin{equation}\label{bottom}
\hspace{-0pt}\left\{
\!\!\!
\begin{array}{l}
\vspace{0.2em}
\hspace{-0.25em}\Big( 1 - {\textstyle\frac{\sigma+1}{2}}\beta_2^{\ast} \nabla h \cdot \!\nabla \Big) \overline{\phi_2} \ms \left( 1 - \alpha_2^{\ast} \Delta \right) \phi_2^{\ast} - \beta_2^{\ast} w_2^{\ast} + \nabla h \cdot\! [ \delta_2^{\ast} \nabla w_2^{\ast} - \varepsilon_2^{\ast} \nabla \phi_2^{\ast} ],\\
\vspace{0.25em}
\hspace{-0.25em}\Big( 1 - {\textstyle\frac{\sigma+1}{2}}\beta_2^{\ast} \nabla h \cdot \!\nabla \Big) \overline{w_2} \ms \beta_2^{\ast} \Delta \phi_2^{\ast} + \left( 1 - \alpha_2^{\ast} \Delta \right) w_2^{\ast}
+ \nabla h \cdot\! [ \gamma_2^{\ast} \nabla w_2^{\ast} ].
\end{array}
\right.
\end{equation}

\subsection{Formulation of the approximate static Dirichlet-Neumann operator}

The final step in deriving our approximate static Dirichlet-Neumann operator $\mathcal{G}_0^{app}[h]$ is to reformulate the boundary conditions at $z=\hat{z}$ and $z=\bar{z}$ in terms of $P_i \phi_i$ and $P_i w_i$ instead of $\phi_i$ and $w_i$. At the interface, the two operators acting on $\widehat{\phi_1}$ and $\widehat{w_1}$ on one side, and $\widehat{\phi_2}$ and $\widehat{w_2}$ on the other side, differ from each other. A simple reformulation of the continuity condition $\widehat{\phi_1} = \widehat{\phi_2}$ is
\begin{eqnarray*}
\left( 1 - \frac{\sigma}{2}\beta_1^{\ast} \nabla h \cdot \nabla \right) \widehat{\phi_1} & \hspace{-0.5em}\ms & \hspace{-0.5em}\Big( 1 - \frac{h}{4} \nabla h \cdot \nabla \Big) \Big( 1 + \frac{\,\sigma+1}{2}\,\beta_2^{\ast} \nabla h \cdot \nabla \Big) \widehat{\phi_2}\;.
\end{eqnarray*}
Consequently, the continuity conditions (\ref{fin5}), (\ref{fin6}) take the form
\begin{equation}\label{inter}
\left\{
\begin{array}{l}
\vspace{0.15em}
\hspace{-0.5em}\left( 1 - \frac{\sigma}{2}\beta_1^{\ast} \nabla h \cdot \nabla \right) \widehat{\phi_1} = \Big( 1 - \frac{h}{4} \nabla h \cdot \nabla \Big) \Big( 1 + \frac{\,\sigma+1}{2}\beta_2^{\ast} \nabla h \cdot \nabla \Big) \widehat{\phi_2}\;,\\
\hspace{-0.5em}\left( 1 - \frac{\sigma}{2}\beta_1^{\ast} \nabla h \cdot \nabla \right) \widehat{w_1} = \Big( 1 - \frac{h}{4} \nabla h \cdot \nabla \Big) \Big( 1 + \frac{\,\sigma+1}{2}\beta_2^{\ast} \nabla h \cdot \nabla \Big) \widehat{w_2}\;.
\end{array}
\right.
\end{equation}
Finally, applying the operator $1 - ((\sigma+1)/2)\beta_2^{\ast} \nabla h \cdot \nabla$ to (\ref{fin7}) leads to the following reformulation of the kinematic condition at the bottom
\begin{equation}\label{seabed}
\Big( 1 - \frac{\sigma+1}{2}\beta_2^{\ast} \nabla h \cdot \nabla \Big) \overline{w_2} + \nabla h \cdot \nabla \Big( 1 - \frac{\sigma+1}{2}\beta_2^{\ast} \nabla h \cdot \nabla \Big) \overline{\phi_2} \hspace{0.25em}\ms\hspace{0.25em} 0\;.
\end{equation}

Gathering all the previous results, we are able to construct the system of five equations on the six unknowns $\phi_0$, $w_0$, $\phi_1^{\ast}$, $w_1^{\ast}$, $\phi_2^{\ast}$, and $w_2^{\ast}$ that defines the approximate operator $\mathcal{G}_0^{app}[h]$ linking $w_0$ to $\phi_0$. The first one corresponds to the reformulated Dirichlet condition $(P_1 \phi_1)|_{z=0} = (P_1|_{z=0}) \phi_0$, i.e. the first equation of (\ref{surf}). The second and third equations correspond to the continuity conditions at the interface $z=\hat{z}$, and are obtained by plugging (\ref{interf}) into (\ref{inter}). The fourth one is the condition at the sea bottom derived by plugging (\ref{bottom}) into (\ref{seabed}), and the last one corresponds to the Neumann condition $(P_1 w_1)|_{z=0} = (P_1|_{z=0}) w_0$ as expressed by the second equation of (\ref{surf}). These five equations can be recast as
\begin{equation}\label{toto}
\left\{
\begin{array}{l}
\vspace{0.5em}
\left(
\begin{array}{cccc}
\vspace{0.35em}
\mathcal{M}_{11} & \mathcal{M}_{12} & 0 & 0\\
\vspace{0.35em}
\mathcal{M}_{21} & \mathcal{M}'_{22} & \mathcal{M}'_{23} & \mathcal{M}'_{24}\\
\vspace{0.35em}
\mathcal{M}_{31} & \mathcal{M}'_{32} & \mathcal{M}'_{33} & \mathcal{M}'_{34}\\
0 & 0 & \mathcal{M}'_{43} & \mathcal{M}'_{44}\\
\end{array}
\right)
\left(
\begin{array}{c}
\vspace{0.35em}
\phi_1^{\ast}\\
\vspace{0.35em}
w_1^{\ast}\\
\vspace{0.35em}
\phi_2^{\ast}\\
w_2^{\ast}\\
\end{array}
\right)
=
\left(
\begin{array}{c}
\vspace{0.35em}
P_0 \phi_0\\
\vspace{0.35em}
0\\
\vspace{0.35em}
0\\
0\\
\end{array}
\right)\;,\\
P_0 {w_0} = - {\beta_1^{\ast}} \Delta \phi_1^{\ast} + \left( 1 - {\alpha_1^{\ast}} \Delta
- {\varepsilon_1^{\ast}} \nabla h \cdot \nabla \right) w_1^{\ast}\;,
\end{array}
\right.
\end{equation}
with the differential operators
\begin{equation}\label{opM}
\hspace{-0em}
\left\{
\begin{array}{ll}
P_0 = 1 + \frac{\sigma}{2}\beta_1^{\ast} \nabla h \cdot \nabla, & \\
\vspace{0.15em}
\mathcal{M}_{11} = 1 - \alpha_1^{\ast} \Delta + \gamma_1^{\ast} \nabla h \cdot \nabla\,, &
\mathcal{M}_{12} = \beta_1^{\ast} - {\delta_1^{\ast}} \nabla h \cdot \nabla\,,\\
\vspace{0.15em}
\mathcal{M}_{21} = 1 - \alpha_1^{\ast} \Delta - \varepsilon_1^{\ast} \nabla h \cdot \nabla\,, &
\mathcal{M}'_{22}= - \beta_1^{\ast} + \delta_1^{\ast} \nabla h \cdot \nabla\,,\\
\vspace{0.15em}
\mathcal{M}'_{23} = - (1 - \alpha_2^{\ast} \Delta) - \nabla h \cdot \Big( (\gamma_2^{\ast} - \frac{h}{4})\nabla + \frac{h}{4}\alpha_2^{\ast}\nabla\Delta \Big)\,,\hspace{-5cm}&\\
\vspace{0.15em}
\mathcal{M}'_{24}=- \beta_2^{\ast} + (\delta_2^{\ast} + \frac{h}{4}\beta_2^{\ast}) \nabla h \cdot \nabla\,,&
\mathcal{M}_{31} = \beta_1^{\ast} \Delta\,,\\
\vspace{0.15em}
\mathcal{M}'_{32}= 1 - \alpha_1^{\ast} \Delta + \gamma_1^{\ast} \nabla h \cdot \nabla\,, &
\mathcal{M}'_{33} = \beta_2^{\ast} \Delta - \frac{h}{4}\beta_2^{\ast} \nabla h \cdot \nabla\Delta\,,\\
\vspace{0.15em}
\mathcal{M}'_{34} = - (1 - \alpha_2^{\ast} \Delta) + \nabla h \cdot \Big( (\varepsilon_2^{\ast}+\frac{h}{4}) \nabla - \frac{h}{4} \alpha_2^{\ast} \nabla\Delta \Big)\,,\hspace{-5cm}&\\
\mathcal{M}'_{43} = \beta_2^{\ast} \Delta + \nabla h \cdot (\nabla - \alpha_2^{\ast} \nabla\Delta)\,, &
\mathcal{M}'_{44} = 1 - \alpha_2^{\ast} \Delta + (\gamma_2^{\ast} - \beta_2^{\ast}) \nabla h \cdot  \nabla\,.\hspace{-3em}
\end{array}
\right.
\end{equation}
Incidentally, we observe that it is possible to eliminate all the third-order derivatives from the operators $\mathcal{M}'_{23}$, $\mathcal{M}'_{33}$, $\mathcal{M}'_{34}$, and $\mathcal{M}'_{43}$ in these equations. Indeed, applying the operator $\nabla h \cdot \nabla$ to respectively the fourth, first, and second equation of (\ref{toto}), using the mild-slope approximation and combining the results yields
\begin{equation}\label{astuce}
\begin{array}{c}
\vspace{0.15em}
\alpha_2^{\ast} \nabla h \cdot \nabla \Delta w_2^{\ast} \ms \nabla h \cdot \nabla w_2^{\ast} + 
\beta_2^{\ast} \nabla h \cdot \nabla \Delta \phi_2^{\ast}\;,\\
\vspace{0.15em}
\alpha_1^{\ast} \nabla h \cdot \nabla \Delta \phi_1^{\ast} \ms \nabla h \cdot \nabla \phi_1^{\ast} + \beta_1^{\ast} \nabla h \cdot \nabla w_1^{\ast} - \nabla h \cdot \nabla \phi_0\;,\\
\alpha_2^{\ast} \nabla h \cdot \nabla \Delta \phi_2^{\ast} \ms 2\beta_1^{\ast} \nabla h \cdot \nabla w_1^{\ast} + \nabla h \cdot \nabla \phi_2^{\ast} +
\beta_2^{\ast} \nabla h \cdot \nabla w_2^{\ast} - \nabla h \cdot \nabla \phi_0\;.
\end{array}
\end{equation}

Plugging these results into (\ref{toto}), we formulate our approximate static Dirichlet-Neumann operator $\mathcal{G}_0^{app}[h]$ as follows:
\begin{equation}\label{TDN}
\left\{
\begin{array}{l}

\vspace{0.5em}

\left(
\begin{array}{cccc}
\vspace{0.35em}
{\mathcal{M}_{11}} & {\mathcal{M}_{12}} & 0 & 0\\
\vspace{0.35em}
\mathcal{M}_{21} & \mathcal{M}_{22} & \mathcal{M}_{23} & \mathcal{M}_{24}\\
\vspace{0.35em}
\mathcal{M}_{31} & \mathcal{M}_{32} & \mathcal{M}_{33} & \mathcal{M}_{34}\\
0 & \mathcal{M}_{42} & \mathcal{M}_{43} & \mathcal{M}_{44}\\
\end{array}
\right)

\left(
\begin{array}{c}
\vspace{0.35em}
\phi_1^{\ast}\\
\vspace{0.35em}
w_1^{\ast}\\
\vspace{0.35em}
\phi_2^{\ast}\\
w_2^{\ast}\\
\end{array}
\right)

=

\left(
\begin{array}{c}
\vspace{0.35em}
P_0 \phi_0\\
\vspace{0.35em}
Q_1 \phi_0\\
\vspace{0.35em}
Q_2 \phi_0\\
Q_3 \phi_0\\
\end{array}
\right)\;,\\

P_0 {w_0} = - {\beta_1^{\ast}} \Delta \phi_1^{\ast} + \left( 1 - {\alpha_1^{\ast}} \Delta 
- {\varepsilon_1^{\ast}} \nabla h \cdot \nabla \right) w_1^{\ast}\;,

\end{array}
\right.
\end{equation}
where the differential operators $P_0$, $\mathcal{M}_{11}$, $\mathcal{M}_{12}$, $\mathcal{M}_{21}$, and $\mathcal{M}_{31}$ are given by (\ref{opM}) and the new operators are defined by
\begin{equation} \label{M}
\hspace{-0.2em}\left\{\hspace{-0.3em}
\begin{array}{lcl}
\mathcal{M}_{22} = - \beta_1^{\ast} + (\delta_1^{\ast} - \frac{h}{2}\beta_1^{\ast}) \nabla h \cdot \nabla, & &
\hspace{-1.25em}\mathcal{M}_{23} = -(1 - \alpha_2^{\ast} \Delta) - \gamma_2^{\ast} \nabla h \cdot \nabla,\\
\mathcal{M}_{24} = - \beta_2^{\ast} + \delta_2^{\ast} \nabla h \cdot \nabla, & &
\hspace{-1.25em}\mathcal{M}_{32} = 1 - \alpha_1^{\ast} \Delta + (\gamma_1^{\ast} - \frac{3\,\sigma}{1-\sigma} h) \nabla h \cdot \nabla,\\
\mathcal{M}_{33} = \beta_2^{\ast} \Delta - \frac{3}{1-\sigma} \nabla h \cdot \nabla, & &
\hspace{-1.25em}\mathcal{M}_{34} = -(1 - \alpha_2^{\ast} \Delta) + (\varepsilon_2^{\ast} - \frac{3}{2}h) \nabla h \cdot \nabla,\\
\mathcal{M}_{42} = -2\beta_1^{\ast} \nabla h \cdot \nabla,\;\mathcal{M}_{43} = \beta_2^{\ast} \Delta, & &
\hspace{-1.25em}\mathcal{M}_{44} = 1 - \alpha_2^{\ast} \Delta + (\gamma_2^{\ast} - 2\beta_2^{\ast}) \nabla h \cdot \nabla,\\
Q_1 = -\frac{h}{4}\nabla h \cdot \nabla\;,\;\;Q_2 = \frac{3}{\sigma-1}\nabla h \cdot \nabla\;,\;\;Q_3 = -\nabla h \cdot \nabla\,.\hspace{-5cm}&\\
\end{array}
\right.
\end{equation}

We denote by $\mathcal{M} = (\mathcal{M}_{ij})_{1 \le i,j \le 4}$ the matrix differential operator linking $\phi_1^{\ast}$, $w_1^{\ast}$, $\phi_2^{\ast}$, and $w_2^{\ast}$ to $\phi_0$ in (\ref{TDN}). We can write each of the operators $\mathcal{M}_{ij}$ as a sum of a first-order (in $h$) operator $\mathcal{P}_{ij}$ and a mild-slope operator $\nabla h \cdot \mathcal{Q}_{ij}$, yielding
\begin{equation}\label{decomp}
\mathcal{M} = \mathcal{P} + \nabla h \cdot \mathcal{Q}\;,
\end{equation}
where
$$
\mathcal{P}
=
\left(
\begin{array}{cccc}
\vspace{0.15em}
1 - {\alpha_1^{\ast}} \Delta & {\beta_1^{\ast}} & 0 & 0 \\
\vspace{0.15em}
1 - \alpha_1^{\ast} \Delta & - \beta_1^{\ast} & -1 + \alpha_2^{\ast}\Delta & - \beta_2^{\ast} \\
\vspace{0.15em}
\beta_1^{\ast} \Delta & 1 - \alpha_1^{\ast} \Delta & \beta_2^{\ast} \Delta & -1 + \alpha_2^{\ast} \Delta \\
0 & 0 & \beta_2^{\ast} \Delta & 1 - \alpha_2^{\ast} \Delta
\end{array}
\right)\;,
$$
$$
\mathcal{Q}
=
\left(
\begin{array}{cccc}
\vspace{0.15em}
\gamma_1^{\ast} \nabla & - \delta_1^{\ast} \nabla & 0 & 0\\
\vspace{0.15em}
- \varepsilon_1^{\ast} \nabla & (\delta_1^{\ast} - {\textstyle\frac{h}{2}}\beta_1^{\ast}) \nabla & - \gamma_2^{\ast} \nabla & \delta_2^{\ast} \nabla\\
\vspace{0.15em}
0 & (\gamma_1^{\ast} - {\textstyle\frac{3\,\sigma}{1-\sigma}} h) \nabla & - {\textstyle\frac{3}{1-\sigma}} \nabla & (\varepsilon_2^{\ast} - {\textstyle\frac{3}{2}}h) \nabla\\
0 & -2\beta_1^{\ast} \nabla & 0 & (\gamma_2^{\ast} - 2\beta_2^{\ast}) \nabla
\end{array}
\hspace{-0.1em}\right)\;\;.
$$ 
We denote by $U$ the vector $(\phi_1^{\ast}, w_1^{\ast}, \phi_2^{\ast}, w_2^{\ast})^T$ and by $F$ the right-hand side \\$(P_0\phi_0, Q_1\phi_0, Q_2\phi_0, Q_3\phi_0)^T$. The differential system in (\ref{TDN}) then takes the form
\begin{equation}\label{sys}
(\mathcal{P} + \nabla h \cdot \mathcal{Q})\,U = F\;.
\end{equation}
A Fourier analysis of the differential operator $\mathcal{P}$ shows that it can be inverted in the case of a flat bottom. Considering an uneven bottom, we can write $\mathcal{P} = \mathcal{P}_0 + \mathcal{P}_H$ where $\mathcal{P}_0 = \mathcal{P}(h_0)$, $h_0$ being the mean depth, and where $\mathcal{P}_H$ is of order $O(\nabla h)$. This ensures that the differential operator $\mathcal{P}$ is invertible for $\nabla h$ small enough. Finally, the differential operator $\mathcal{R}$ defined by
\begin{equation}\label{R}
\mathcal{R} = \mathcal{P}^{-1}(Id - \nabla h \cdot \mathcal{Q} \mathcal{P}^{-1})\;,
\end{equation}
where $Id$ is the identity operator, is thus an approximate inverse of the operator $\mathcal{P} + \nabla h \cdot
\mathcal{Q}$ up to $O(|\nabla h|^2)$ terms. Hence,
\begin{equation}\label{sys3}
U \ms \mathcal{R} F\;,
\end{equation}
which yields the explicit expressions of $\phi_1^{\ast}$, $w_1^{\ast}$, $\phi_2^{\ast}$, and $w_2^{\ast}$ in terms of $\phi_0$.\\  
\noindent The very last step consists in introducing the operators
\begin{equation}\label{G0app2}
\left\{
\begin{array}{l}
\mathcal{N}_1 = - {\beta_1^{\ast}} \Delta\;,\\
\mathcal{N}_2 = 1 - {\alpha_1^{\ast}} \Delta - {\varepsilon_1^{\ast}} \nabla h \cdot \nabla\;,\\
Q_0 = 1 - \frac{\sigma}{2}\beta_1^{\ast} \nabla h \cdot \nabla \ms P_0^{-1}\;,
\end{array}
\right.
\end{equation}
and plugging the expressions of $\phi_1^{\ast}$ and $w_1^{\ast}$ into the last equation of (\ref{TDN}), so as to obtain the explicit relation between $w_0$ and $\phi_0$, and thus the explicit expression of our approximate static Dirichlet-Neumann operator $\mathcal{G}_0^{app\,}[h]$
\begin{equation}\label{G0app1}
\mathcal{G}_0^{app\,}[h] =
Q_0\
\left(
\begin{array}{cccc}
\vspace{0.15em}
\mathcal{N}_1 & 0 & 0 & 0\\
\vspace{0.15em}
0& \mathcal{N}_2 & 0 & 0\\
\vspace{0.15em}
0 & 0 & 0 & 0\\
0 & 0 & 0 & 0
\end{array}
\right)
\mathcal{R}
\left(\begin{array}{c}\vspace{0.15em}P_0\\\vspace{0.15em}Q_1\\\vspace{0.15em}Q_2\\Q_3\end{array}\right)\,.
\end{equation}
This expression completes the formulation (\ref{model}) of our double-layer Boussinesq-type model and will be further improved in \S 4$\,f$ to tighten the model shoaling properties. Once again, we stress that the major advantage of the operator $\mathcal{G}_0^{app\,}[h]$ is that it is static. Hence, we can construct it at $t=0$ once and for all.\\ 

Once we have computed $\phi_0$, we can compute the values for $\phi_1^{\ast}$, $\phi_2^{\ast}$, and $w_2^{\ast}$ using (\ref{sys3}). Therefore, we can recover the vertical profiles of the velocity potentials $\phi_1$, $\phi_2$ and the vertical velocities $w_1$, $w_2$ over the whole water column using a generalization of (\ref{closure}) for any $z \in (0,\eta)$, and plugging the computed values of $p_i$ and $q_i$ into (\ref{padé}). We point out that we have neglected the third- and fourth-order derivatives in the following expressions, to obtain only second-order derivatives. Of course, there is a price to pay for this choice, as discussed in the linear analysis of the vertical profiles in \S 4$\,e$. We use the expressions
\vspace*{-0.3em}
\begin{equation}\label{profils1}
\left\{
\begin{array}{rcl}
\hspace{-0.25em}\phi_1(t,X,z) & \hspace{-0.5em}= & \hspace{-0.5em}(1-\frac{z^2}{2}\Delta)\, \phi_0 + (z-\frac{z^3}{6}\Delta)\, w_0\;,\\
\hspace{-0.25em}w_1(t,X,z) & \hspace{-0.5em}= & \hspace{-0.5em}-z \Delta\, \phi_0 + (1-\frac{z^2}{2}\Delta)\, w_0\;,
\end{array}
\right.
\end{equation}
in the vertical region $z \in (0,\eta)$ and the expressions
\begin{equation}\label{profils2}
\hspace{-0em}
\left\{
\!\!\!
\begin{array}{l}
\phi_i(t,X,z) \ms (1 + \beta_i\nabla z_i\cdot\nabla) \left[ \left(1 - \alpha_i^{\ddag}(z) \Delta \right) \phi_i^{\ast} + \left(\beta_i^{\ddag}(z) - \gamma_i^{\ddag}(z) \Delta \right) w_i^{\ast}\right.\\
\left. \hspace{12.99em} - \beta_i^{\ddag}(z) \nabla z_i \cdot \nabla \phi_i^{\ast} \right] \;,\\
w_i(t,X,z) \ms (1 + \beta_i\nabla z_i\cdot\nabla) \left[ -\beta_i^{\ddag}(z) \Delta \phi_i^{\ast} + \left(1 - \alpha_i^{\ddag}(z) \Delta \right) w_i^{\ast}\right.\\
\left. \hspace{12.99em} + \beta_i^{\ddag}(z) \nabla z_i \cdot \nabla w_i^{\ast} \right]\;,
\end{array}
\right.
\end{equation}
for $z \in [\hat{z}, \min(0,\eta)]$ if $i=1$ and for $z \in [\bar{z}, \hat{z}]$ if $i=2$, where
\begin{equation}\label{coeffsprofile}
\hspace{-0em}
\left\{
\begin{array}{ll}
\hspace{-0.35em}\alpha_1^{\ddag}(z) = \frac{z}{2} (z+\sigma h) + \frac{\sigma^2}{12} h^2\,, &
\alpha_2^{\ddag}(z) = \frac{1}{2} (z+h) (z+\sigma h) + \frac{(1-\sigma)^2}{12} h^2\,,\\
\hspace{-0.35em}\beta_1^{\ddag}(z) = z + \frac{\sigma}{2} h\,, &
\beta_2^{\ddag}(z) = z + \frac{\sigma+1}{2} h\,,\\
\hspace{-0.35em}\gamma_1^{\ddag}(z) = \frac{z}{6} (z+\frac{\sigma}{2} h)(z + \sigma h)\,, &
\gamma_2^{\ddag}(z) = \frac{1}{6} (z+h) (z+\frac{\sigma+1}{2} h)(z + \sigma h)\,.
\end{array}\hspace{-2em}
\right.
\end{equation}
Using (\ref{profils1}) in the region between the still-water level and the free surface and (\ref{profils2}) elsewhere instead of applying (\ref{profils2}) everywhere seems to provide a more accurate description of the nonlinear profiles, as specified by Madsen \textit{et al.} (2002, 2003). This property has also been observed during the nonlinear simulations performed on the present model in \S 5.

\section{Linear analysis of the double-layer model}

The goal of this section is to analyze the linear properties of the model (namely the phase and group velocities, the vertical profiles of velocity potential and vertical velocity, and the linear shoaling) and to optimize their accuracy in relation to the results of Stokes linear theory.

\subsection{Linearization of the governing equations}

In order to investigate these linear properties, we restrict the analysis to the one-dimensional case. We linearize the governing equations (\ref{model}) around steady-state, which yields $\widetilde{\phi_1} = \phi_0$ and $\widetilde{w_1} = w_0$, and leads to the linearized model
\begin{subnumcases}{}
\partial_t \phi_0 + g\eta = 0\;,\label{lin1a}\\
\partial_t \eta - w_0 = 0\;,\label{lin2a}\\
w_0 = \mathcal{G}_0^{app\,}[h] \phi_0\;,\label{lin3a}
\end{subnumcases}
or equivalently, up to $O(h_x^2)$ terms,
\begin{subnumcases}{}
\partial_t \phi_0 + g\eta = 0\;,\label{lin1}\\
\partial_t \eta - w_0 = 0\;,\label{lin2}\\
\left(
\begin{array}{cccc}
\vspace{0.25em}
{\mathcal{M}_{11}} & {\mathcal{M}_{12}} & 0 & 0\\
\vspace{0.25em}
\mathcal{M}_{21} & \mathcal{M}_{22} & \mathcal{M}_{23} & \mathcal{M}_{24}\\
\vspace{0.25em}
\mathcal{M}_{31} & \mathcal{M}_{32} & \mathcal{M}_{33} & \mathcal{M}_{34}\\
0 & \mathcal{M}_{42} & \mathcal{M}_{43} & \mathcal{M}_{44}\\
\end{array}
\right)
\left(
\begin{array}{c}
\vspace{0.25em}
\phi_1^{\ast}\\
\vspace{0.25em}
w_1^{\ast}\\
\vspace{0.25em}
\phi_2^{\ast}\\
w_2^{\ast}\\
\end{array}
\right)
=
\left(
\begin{array}{c}
\vspace{0.25em}
P_0 \phi_0\\
\vspace{0.25em}
Q_1 \phi_0\\
\vspace{0.25em}
Q_2 \phi_0\\
Q_3 \phi_0\\
\end{array}
\right)\;,\label{lin3}\\
P_0 {w_0} = - {\beta_1^{\ast}} \partial_x^2 \phi_1^{\ast} + \left( 1 - {\alpha_1^{\ast}} \partial_x^2 
- {\varepsilon_1^{\ast}} h_x \partial_x \right) w_1^{\ast}\;,\label{lin4}
\end{subnumcases}
where $h_x$ is the bottom slope and the differential operators $\mathcal{M}_{ij}$ are left unchanged, except that they are here 1-D operators. We point out that in this linearized model, the slope terms are kept in order to investigate the linear shoaling properties. For convenience, we apply the operator $P_0$ to equations (\ref{lin1}), (\ref{lin2}) to obtain
\begin{subnumcases}{}
\partial_t \phi^0 + g N = 0\;,\label{lin21}\\
\partial_t N - W^0 = 0\;,\label{lin22}\\
\left(
\begin{array}{cccc}
\vspace{0.25em}
{\mathcal{M}_{11}} & {\mathcal{M}_{12}} & 0 & 0\\
\vspace{0.25em}
\mathcal{M}_{21} & \mathcal{M}_{22} & \mathcal{M}_{23} & \mathcal{M}_{24}\\
\vspace{0.25em}
\mathcal{M}_{31} & \mathcal{M}_{32} & \mathcal{M}_{33} & \mathcal{M}_{34}\\
0 & \mathcal{M}_{42} & \mathcal{M}_{43} & \mathcal{M}_{44}\\
\end{array}
\right)
\left(
\begin{array}{c}
\vspace{0.25em}
\phi_1^{\ast}\\
\vspace{0.25em}
w_1^{\ast}\\
\vspace{0.25em}
\phi_2^{\ast}\\
w_2^{\ast}\\
\end{array}
\right)
=
\left(
\begin{array}{c}
\vspace{0.25em}
\phi^0\\
\vspace{0.25em}
Q_1 \phi^0\\
\vspace{0.25em}
Q_2 \phi^0\\
Q_3 \phi^0\\
\end{array}
\right)\;,\label{lin23}\\
W_0 = - {\beta_1^{\ast}} \partial_x^2 \phi_1^{\ast} + \left( 1 - {\alpha_1^{\ast}} \partial_x^2 
- {\varepsilon_1^{\ast}} h_x \partial_x \right) w_1^{\ast}\;,\label{lin24}
\end{subnumcases}
where $\phi^0 = P_0 \phi_0, N = P_0 \eta$, $W_0 = P_0 w_0$, and where we have plugged the relation $P_0^{-1} \ms 1 - ({\sigma}/{2})\beta_1^{\ast} h_x \partial_x$ into $\phi_0 = P_0^{-1}\phi^0$ to obtain $Q_i \phi_0 = Q_i P_0^{-1} \phi^0 \ms Q_i \phi^0$ for $i \in \{1,3\}$, since $Q_i$ are differential operators of order $O(h_x)$.\\
Plugging the expression of $\phi^0$ and $W^0$ in terms of $\phi_1^{\ast}$ and $w_1^{\ast}$ (given respectively by the first line of the differential system (\ref{lin23}) and by (\ref{lin24})) into equations (\ref{lin21}) and (\ref{lin22}) leads to the final reformulation of the linearized model
\begin{equation}\label{linf}
\left\{
\begin{array}{l}
\vspace{0.1em}
\mathcal{M}_{11}\,\partial_t \phi_1^{\ast} + \mathcal{M}_{12}\,\partial_t w_1^{\ast} + g N = 0\;,\\
\vspace{0.1em}
\partial_t N + \left[ \beta_1^{\ast} \partial_x^2 \right] \phi_1^{\ast} - \left[ 1 - \alpha_1^{\ast} \partial_x^2 - \varepsilon_1^{\ast} h_x \partial_x \right] w_1^{\ast} = 0\;,\\
\vspace{0.1em}
[\mathcal{M}_{21} - Q_1 \mathcal{M}_{11}]\,\phi_1^{\ast} + [\mathcal{M}_{22} - Q_1 \mathcal{M}_{12}]\,w_1^{\ast} + \mathcal{M}_{23}\,\phi_2^{\ast} + \mathcal{M}_{24}\,w_2^{\ast} = 0\;,\\
\vspace{0.1em}
[\mathcal{M}_{31} - Q_2 \mathcal{M}_{11}]\,\phi_1^{\ast} + [\mathcal{M}_{32} - Q_2 \mathcal{M}_{12}]\,w_1^{\ast} + \mathcal{M}_{33}\,\phi_2^{\ast} + \mathcal{M}_{34}\,w_2^{\ast} = 0\;,\\
- Q_3 \mathcal{M}_{11}\,\phi_1^{\ast} - Q_3 \mathcal{M}_{12}\,w_1^{\ast} + \mathcal{M}_{43}\,\phi_2^{\ast} + \mathcal{M}_{44}\,w_2^{\ast} = 0\;.
\end{array}
\right.
\end{equation}

\noindent We now look for solutions of the classical form
\begin{equation}\label{sollin}
\left\{
\begin{array}{l}
\eta(x,t) = A \exp{i\theta}\,,\;\;\theta = \omega t - k x\;,\\
\phi_1^{\ast} = -i (B_1 + i h_x B_2) \exp{i\theta}\,,\\
w_1^{\ast} = i (C_1 + i h_x C_2) \exp{i\theta}\;,\\
\phi_2^{\ast} = -i (D_1 + i h_x D_2) \exp{i\theta}\,,\\
w_2^{\ast} = i (E_1 + i h_x E_2) \exp{i\theta}\;,
\end{array}
\right.
\end{equation}
where $A,B_1,B_2,C_1,C_2,D_1,D_2,E_1$, and $E_2$ are slowly spatially-varying functions (i.e. of the general form $F(\nu x)$ with $\nu \ll 1$), $k$ is the wavenumber, and $\omega$ the wave frequency. The complex conjugate parts of these expressions have been left out for brevity. As stated in Madsen \textit{et al.} (2002, 2003), the $B_2,C_2,D_2$, and $E_2$ contributions are necessary because of the bottom slope, and since the velocity potential variables are not in phase with the free surface at the lowest order in $h_x$, but are so at the next order.

\subsection{Linear dispersion relation}

To determine the linear properties of the two-layer model, we substitute the desired form of solutions (\ref{sollin}) into the linear formulation (\ref{linf}) and collect terms at the lowest order in $h_x$. Thus, we obtain a linear system of five homogeneous equations in $A,B_1,C_1,D_1$, and $E_1$. This system has non-trivial solutions if its determinant vanishes, which yields the following dispersion relation
\begin{equation}\label{disp}
\frac{c^2}{gh} = \frac{\omega^2}{ghk^2} = \frac{1+a_2(kh)^2+a_4(kh)^4+a_6(kh)^6}{1+b_2(kh)^2+b_4(kh)^4+b_6(kh)^6+b_8(kh)^8}\;,
\end{equation}
where $c$ is the wave celerity and where the $(a_i)$ and $(b_i)$ coefficients are given by
$$
\left\{
\begin{array}{l}
\vspace{0.2em}
a_2 = 2S+\frac{1}{12}\;,\;\;
a_4 = S(2S+\frac{1}{12})\;,\;\;
a_6 = S^3\;,\\
b_2 = 2S+\frac{5}{12}\;,\;\;
b_4 = 3S^2+\frac{2}{3}S+\frac{1}{144}\;,\;\;
b_6 = S^2(2S+\frac{5}{12})\;,\;\;
b_8 = S^4\;,
\end{array}
\right.
$$
where $S = \sigma (1-\sigma)/12$. This dispersion relation is compared in \S 4$\,e$ to the exact linear dispersion relation given by Stokes linear theory, namely
\begin{equation}\label{Stokes1}
\frac{c_s^2}{gh} = \frac{\omega^2}{ghk^2} = \frac{\tanh{kh}}{kh}\;,
\end{equation}
and to its $[6,8]$ Padé approximation which has the same rational form as (\ref{disp}).

\subsection{Linear vertical profiles}

Coming back to the previous linear system in $A,B_1,C_1,D_1$, and $E_1$, we can now express each of the unknowns $B_1,C_1,D_1$, and $E_1$ in terms of $A$, which leads to tedious expressions that are not given here for brevity. For a flat bottom, the expressions of $\phi_i$ and $w_i$ on the whole water column are given by (\ref{profils2}), (\ref{coeffsprofile}) since
\begin{equation}\label{padéé}
\left\{
\begin{array}{lll}
\vspace{0.25em}
\hspace{-0.25em}\phi_i(t,X,z) & \hspace{-0.5em}= & \hspace{-0.5em}\left(1 - \alpha_i^{\ddag}(z) \Delta \right) \phi_i^{\ast} + \left(\beta_i^{\ddag}(z) - \gamma_i^{\ddag}(z) \Delta \right) w_i^{\ast}\;,\\
\hspace{-0.25em}w_i(t,X,z) & \hspace{-0.5em}= & \hspace{-0.5em}-\beta_i^{\ddag}(z) \Delta \phi_i^{\ast} + \left(1 - \alpha_i^{\ddag}(z) \Delta \right) w_i^{\ast}\;.
\end{array}
\right.
\end{equation}
Plugging the ansatz (\ref{sollin}) for a flat bottom (i.e. without the mild-slope contributions) into the previous expressions and using the computed values of $B_1,C_1,D_1$, and $E_1$ in terms of $A$ leads to the expressions of $\phi_1(z),w_1(z)$ in the upper layer and $\phi_2(z),w_2(z)$ in the lower layer, in terms of $k$, $h$, $\omega$, $A$, and $\sigma$. Finally, we recover the linear vertical profiles over the whole water column using
\begin{equation}\label{prof}
\hspace{-0em}
\begin{array}{lcl}
\phi(z) = 
\left\{
\begin{array}{l}
\phi_1(z) \;\;\text{for $\hat{z} \le z \le 0$}\;,\\
\phi_2(z) \;\;\text{for $\bar{z} \le z \le \hat{z}$}\;,
\end{array}
\right.
& \text{and} &
w(z) = 
\left\{
\begin{array}{l}
w_1(z) \;\;\text{for $\hat{z} \le z \le 0$}\;,\\
w_2(z) \;\;\text{for $\bar{z} \le z \le \hat{z}$}\;.
\end{array}
\right.
\end{array}\hspace{-1em}
\end{equation}
The resulting vertical profiles will be compared in \S 4$\,e$ to the theoretical linear profiles $\phi_s(z)$ and $w_s(z)$ coming from Stokes linear theory, namely
\begin{equation}\label{Stokes2}
\left\{
\begin{array}{l}
\vspace{0.5em}
\phi_s(z) = \frac{Ag}{\omega} \frac{\cosh{k(z+h)}}{\cosh{kh}} \sin(\omega t - kx)\;,\\
w_s(z) = \frac{Agk}{\omega} \frac{\sinh{k(z+h)}}{\cosh{kh}} \sin(\omega t - kx)\;.
\end{array}
\right.
\end{equation}

\subsection{Linear shoaling}

We now aim at determining the linear shoaling gradient $\gamma_0$ of the double-layer model defined by
\begin{equation}\label{gamma0}
\frac{A_x}{A} = \gamma_0 \frac{h_x}{h}\;.
\end{equation}
In order to determine this shoaling gradient, we use the method proposed by Madsen \textit{et al.} (2002, 2003). Coming back to the linear formulation (\ref{linf}) together with the ansatz (\ref{sollin}), we then collect terms at the next order, i.e. terms proportional to the first derivatives of all the variables. Doing this leads to a new inhomogeneous system of linear equations on the unknowns $A_x,B_2,C_2,D_2$, and $E_2$ involving the first derivatives of $k$ and $h$ (namely $k_x$ and $h_x$) and the first derivatives of $B_1,C_1,D_1$, and $E_1$. Differentiating the previously computed expressions of $B_1,C_1,D_1$, and $E_1$ in terms of $A$, the derivatives of $B_1,C_1,D_1$, and $E_1$ can be expressed only in terms of $A_x$, $k_x$, and $h_x$. Then, differentiating the linear dispersion relation (\ref{disp}) allows to express $k_x$ in terms of $k$, $h$, $h_x$, and $\sigma$. Plugging all these relations into the inhomogeneous system on $A_x,B_2,C_2,D_2$, and $E_2$, we are able to eliminate all the unknowns but $A_x$ and express it in terms of $A$, $h_x$, $kh$, and $\sigma$, thereby yielding the linear shoaling gradient $\gamma_0$ of our double-layer model. The detailed analytic expression for $\gamma_0$ is not reported here for brevity.

The computed shoaling gradient will be compared in \S 4$\,f$ to the exact shoaling gradient $\gamma_s$, which was derived by Madsen \& S\o rensen (1992) using energy flux conservation combined with Stokes linear theory, namely
\begin{equation}\label{Stokes3}
\gamma_s = \frac{2kh\sinh{2kh} + 2k^2h^2(1-\cosh{2kh})}{(2kh+\sinh{2kh})^2}\;.
\end{equation} 

\subsection{Optimization of linear properties}

The goal is now to optimize the linear properties of our double-layer model by minimizing the errors between these properties and the exact linear properties coming from Stokes linear theory. To this end, we tune the free parameter $\sigma$ which defines the interface level $\hat{z} = - \sigma h$.\\
The different errors between the model linear properties and the theoretical ones are computed as follows. We respectively measure the errors on the phase celerity, the vertical profiles of the velocity potential and the vertical velocity, and the linear shoaling gradient as
\begin{equation}
\mathcal{E}_{\alpha}(K,\sigma) = \sqrt{\int_0^{K} \frac{1}{kh}\,E_{\alpha}^{2}(\sigma,kh)\,d(kh)}\,,
\end{equation}
with $\alpha \in \{c,\phi,w,\gamma\}$, $K$ being a reference relative water depth, and
\begin{equation}
\left\{
\begin{array}{l}
E_c^2(\sigma,kh) = \left(\frac{c-c_s}{c_s}\right)^2\,,\;\;
E_{\phi}^2(\sigma,kh) = \frac{1}{h} \int_{-h}^{0} \left(\frac{\phi(z)-\phi_s(z)}{\phi_s(0)}\right)^2 dz\,,\\
E_w^2(\sigma,kh) = \frac{1}{h} \int_{-h}^{0} \left(\frac{w(z)-w_s(z)}{w_s(0)}\right)^2 dz\,,\;\;
E_{\gamma}^2(\sigma,kh) = \left(\gamma_0 - \gamma_s\right)^2\,,
\end{array}
\right.
\end{equation}
where $c$, $c_s$, $(\phi,w)$, $(\phi_s,w_s)$, $\gamma_0$, and $\gamma_s$ come respectively from (\ref{disp}), (\ref{Stokes1}), (\ref{prof}), (\ref{Stokes2}), (\ref{gamma0}), and (\ref{Stokes3}).\\
We point out that in all these errors, the weighting by $\tfrac{1}{kh}$ helps keeping the errors to a minimum for low wave numbers, like in Madsen \textit{et al.} (2002, 2003). Doing this, we sacrifice some accuracy at very high wavenumbers (i.e. in deep water), but we reinforce the model accuracy in shallow water. This weighting by $\tfrac{1}{kh}$ is especially well-suited for the shoaling gradient errors, which are far more critical in shallow water than in deep water.

At this point, we could have minimized each of these errors individually, but doing this leads to quite different optimal values for $\sigma$, ranging from $0.2$ to $0.5$. Furthermore, the minimization of the shoaling gradient error is quite problematic: we will see later that whatever value we choose for $\sigma$, the range of validity in $kh$ is limited. We thus choose to minimize $\mathcal{E}_{c},\mathcal{E}_{\phi}$, and $\mathcal{E}_{w}$ simultaneously to infer the optimal value for $\sigma$, and then optimize the shoaling gradient error $\mathcal{E}_{\gamma}$ differently. We start with the minimization of the errors $\mathcal{E}_{c},\mathcal{E}_{\phi}$, and $\mathcal{E}_{w}$ through the average error $\mathcal{E}_{tot}(K,\sigma) = {\textstyle \frac{1}{3}} (\mathcal{E}_{c}(K,\sigma) + \mathcal{E}_{\phi}(K,\sigma) + \mathcal{E}_{w}(K,\sigma))$, and we compute the optimal value of $\sigma$ for several typical values of $K$: the shallow-water value $K=\pi/2$, the intermediate depth value $K=\pi$, and the deep water values $K=2\pi$ and $K=10$. In this work, we do not optimize $\sigma$ for larger values as the vertical profiles have systematically shown an error peak of at least $2\%$ within the range $kh \in [0,10]$ for larger $K$, for instance $K=15$ or $K=20$. Table \ref{table1} summarizes the optimal values $\sigma_{opt}$ for each value of $K$.
\begin{table}
\caption{Optimal values for $\sigma$}
\begin{center}
\begin{tabular}{|c|c|c|c|c|}
\hline
$K$ & $\pi/2$ & $\pi$ & $2\pi$ & $10$\\
\hline
\;\;\;\;$\sigma_{opt}$\;\;\;\; & \;\;\;$0.473$\;\;\; & \;\;\;$0.428$\;\;\; & \;\;\;$0.365$\;\;\; & \;\;\;$0.314$\;\;\;\\
\hline
\end{tabular}
\end{center}
\label{table1}
\end{table}

\vspace{1em}

\begin{figure}
%\begin{minipage}{0.57\textwidth}
%\psfrag{ccs}[c][l][1.05][270]{$\frac{c}{c_s}\textcolor{white}{\,}$}
%\psfrag{kh}[][][0.8][0]{$kh$}
%\centerline{\includegraphics[width=\textwidth]{disp1.eps}}
%\end{minipage}
%\hspace{-30pt}
%\begin{minipage}{0.53\textwidth}
%\psfrag{ccs}[r][l][1.05][270]{$\frac{c}{c_s}$}
%\psfrag{kh}[][][0.8][0]{$kh$}
%\centerline{\includegraphics[width=\textwidth]{disp2.eps}}
%\end{minipage}
%\hspace{0.75em}
%\begin{minipage}{0.98\textwidth}
%\psfrag{ccs}[r][l][1.05][270]{$\frac{c}{c_s}$}
%\psfrag{kh}[][][0.8][0]{$kh$}
%\center{\includegraphics[width=\textwidth]{disp3.eps}}
%\end{minipage}
\hspace{1pt}
\psfrag{ccs}[c][l][0.97][270]{$\frac{c}{c_s}\textcolor{white}{\;}$}
\psfrag{kh}[][][0.8][0]{$kh$}
\includegraphics[width=0.6\textwidth,height=130pt]{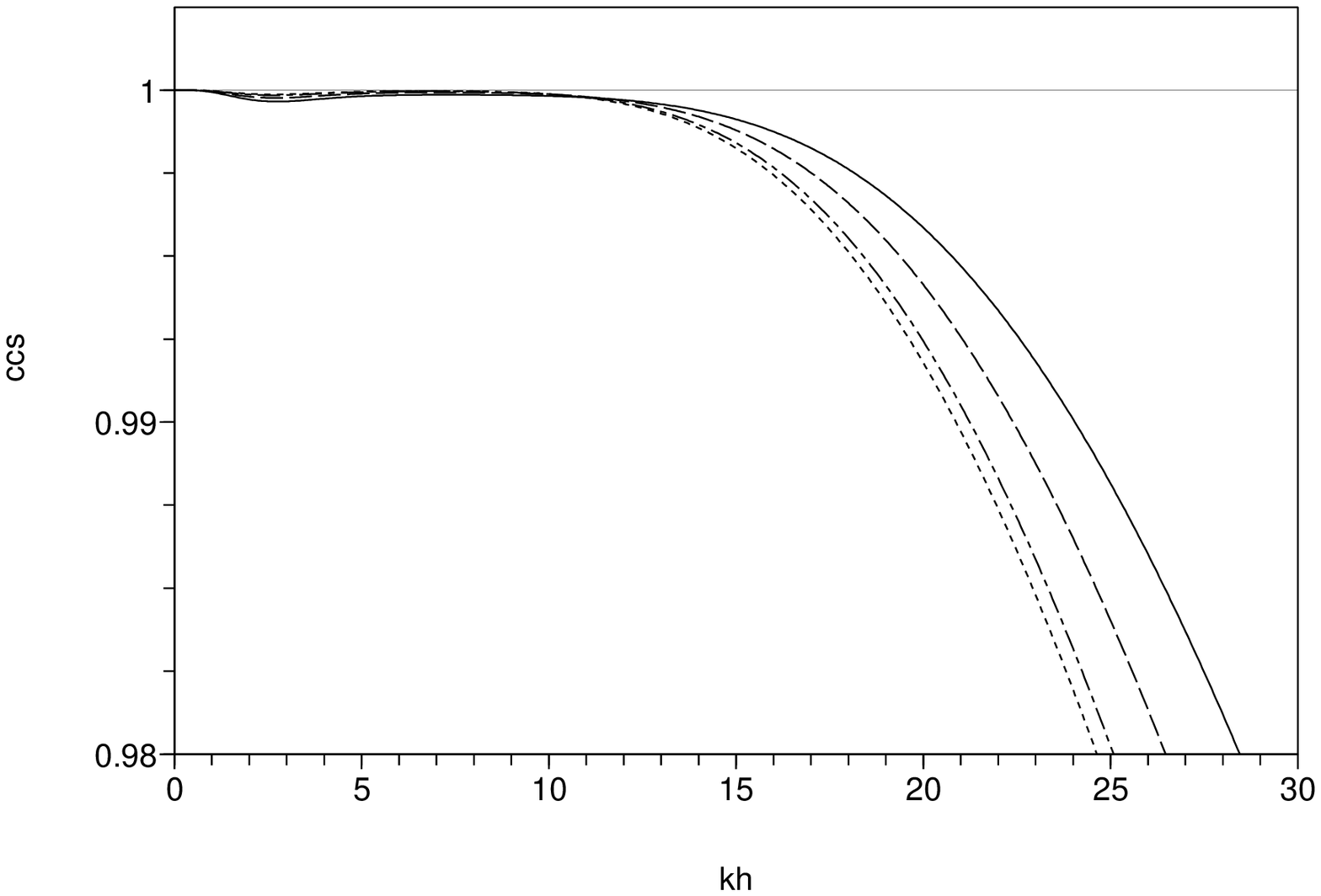}
\psfrag{ccs}[r][l][1.05][270]{$\frac{c}{c_s}$}
\psfrag{kh}[][][0.8][0]{$kh$}
\hspace{-2.5em}
\includegraphics[width=0.43\textwidth,height=130pt]{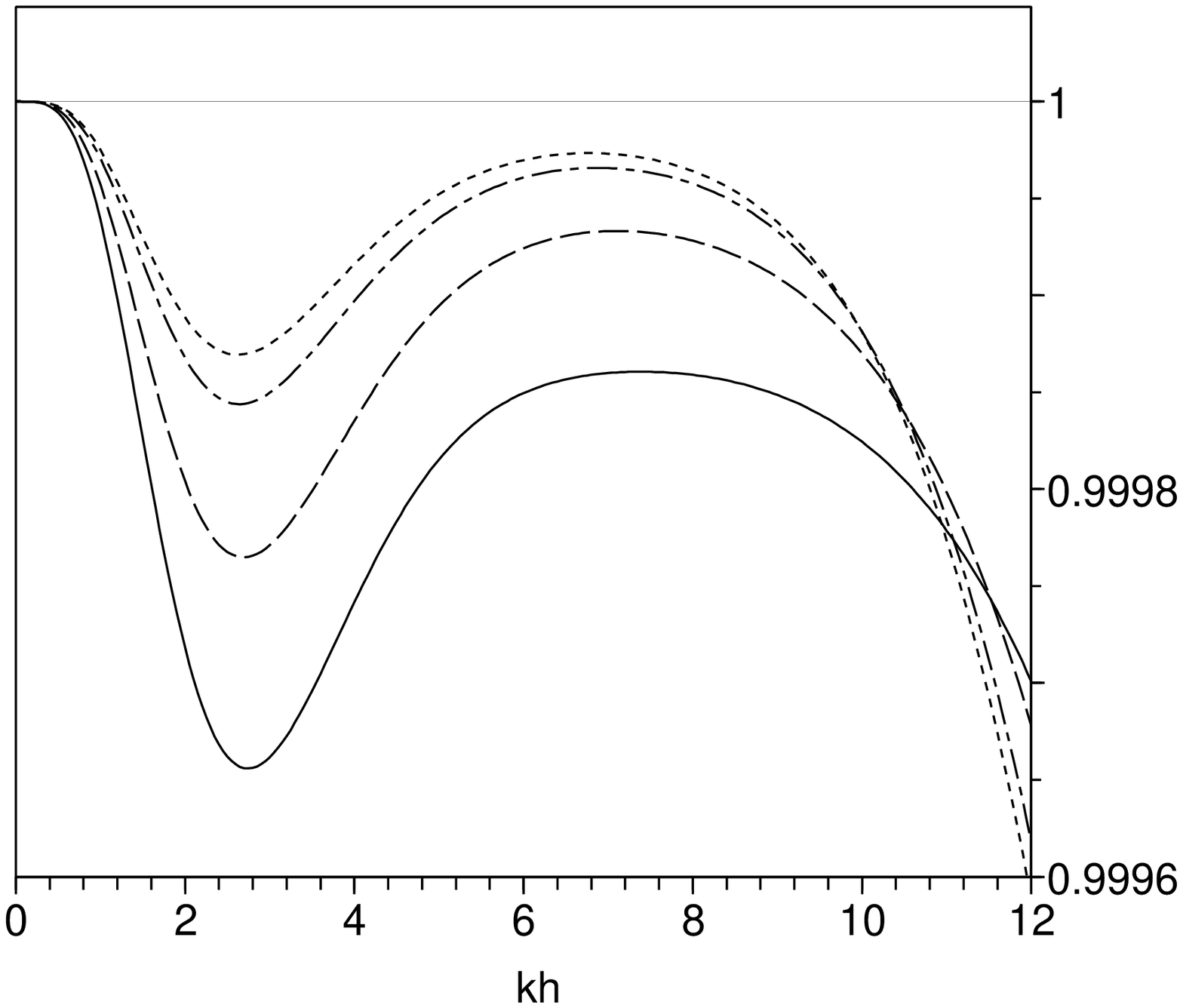}
\psfrag{ccs}[c][l][0.97][270]{$\frac{c}{c_s}\textcolor{white}{\;}$}
\psfrag{kh}[][][0.8][0]{$kh$}
\center{\includegraphics[width=0.6\textwidth,height=130pt]{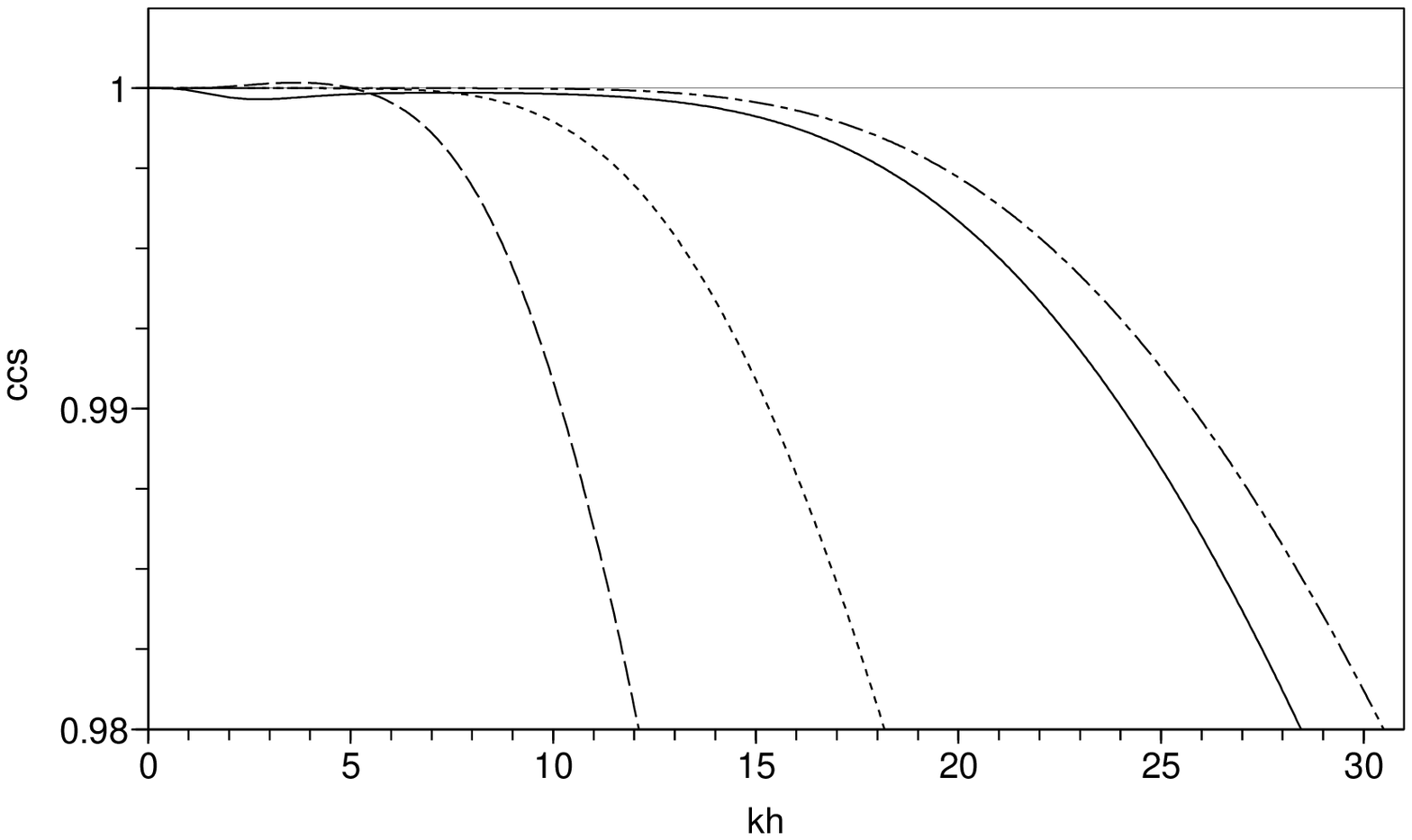}}
\caption{Comparison of linear phase celerity with the exact Stokes result. The top figures compare the errors obtained for our model with $\sigma = 0.314$ (solid line), $\sigma = 0.365$ (dashed line), $\sigma = 0.428$ (dash-dotted line), and $\sigma = 0.473$ (dotted line). The top right figure is a zoom on the region $kh \in [0,12]$. The bottom figure compares the errors for different models: the solid line represents our double-layer model with $\sigma = 0.314$, the dotted line the Padé $[6,8]$ approximation, the dashed line the model of Jamois \textit{et al.} (2006), and the dash-dotted line the model of Madsen \textit{et al.} (2002, 2003).}
\label{figure2}
\end{figure}

Figure \ref{figure2} plots $c/c_s$ to assess the dispersion error on the phase celerity. The upper panel compares the errors obtained for each value of $\sigma_{opt}$. The price to pay for extending the linear range of validity towards deep water values is the growth of a small error peak around $kh = 3$. Indeed, we can see that a $2\%$ error is reached at the very deep water value $kh=24$ for $\sigma_{opt}=0.473$ with a very small $0.01\%$ error peak at $kh \approx 3$, whereas the same error is reached at $kh=28$ for $\sigma_{opt}=0.314$, but with a $0.04\%$ error peak at $kh \approx 3$. However, such an error is not significant, and the overall accuracy of the double-layer model for $\sigma_{opt}=0.314$ appears to be excellent up to very deep water. In the same way, the lower panel of figure \ref{figure2} compares the error on the phase celerity of our double-layer model with $\sigma_{opt}=0.314$ with the errors obtained for the Padé $[6,8]$ approximation, the model of Jamois \textit{et al.} (2006), and the one of Madsen \textit{et al.} (2002, 2003). We remark that our double-layer model accuracy is far better in deep water than what is achieved with the Padé $[6,8]$ approximation and the model of Jamois \textit{et al.} (2006): a $2\%$ error is reached at the very deep water value $kh \approx 28$ for the double-layer model, whereas the same error is already reached at $kh \approx 18$ for the Padé $[6,8]$ approximation and $kh \approx 12$ for the model of Jamois \textit{et al.} (2006). In comparison with these results, Lynett \& Liu (2004$a$, $b$) showed that their double-layer model reaches the same $2\%$ error (not plotted here) at $kh = 8$. As for the model derived by Madsen \textit{et al.} (2002, 2003), a $2\%$ error is reached at $kh = 30$, i.e. at a slightly greater value than the one reached by our model with $\sigma=0.314$.

\begin{figure}
\psfrag{Ephi}[][][1][270]{$E_{\phi}$}
\psfrag{kh}[][][0.8][0]{$kh$}
\center{\epsfig{file=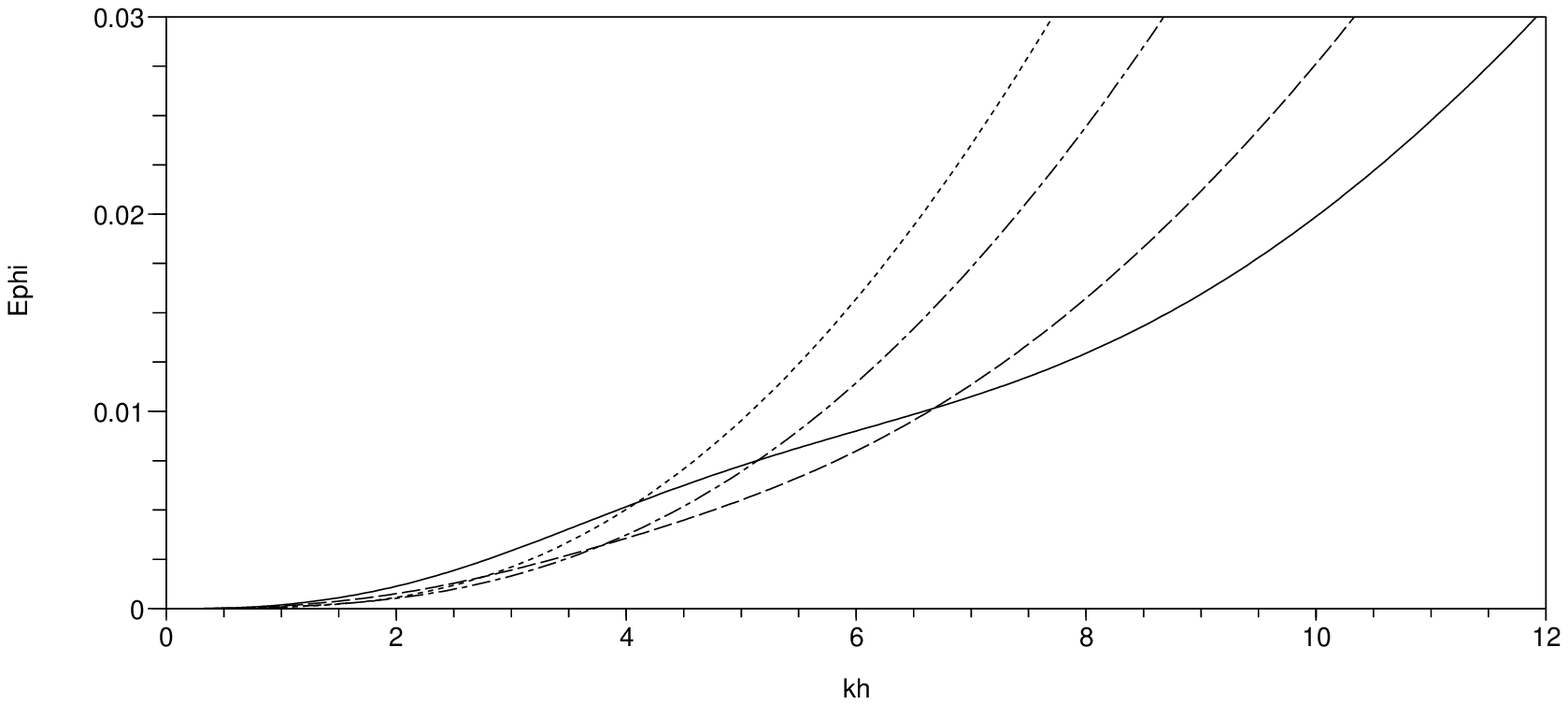,width=\linewidth}}
\psfrag{Ew}[][][1][270]{$E_{w}$}
\psfrag{kh}[][][0.8][0]{$kh$}
\center{\epsfig{file=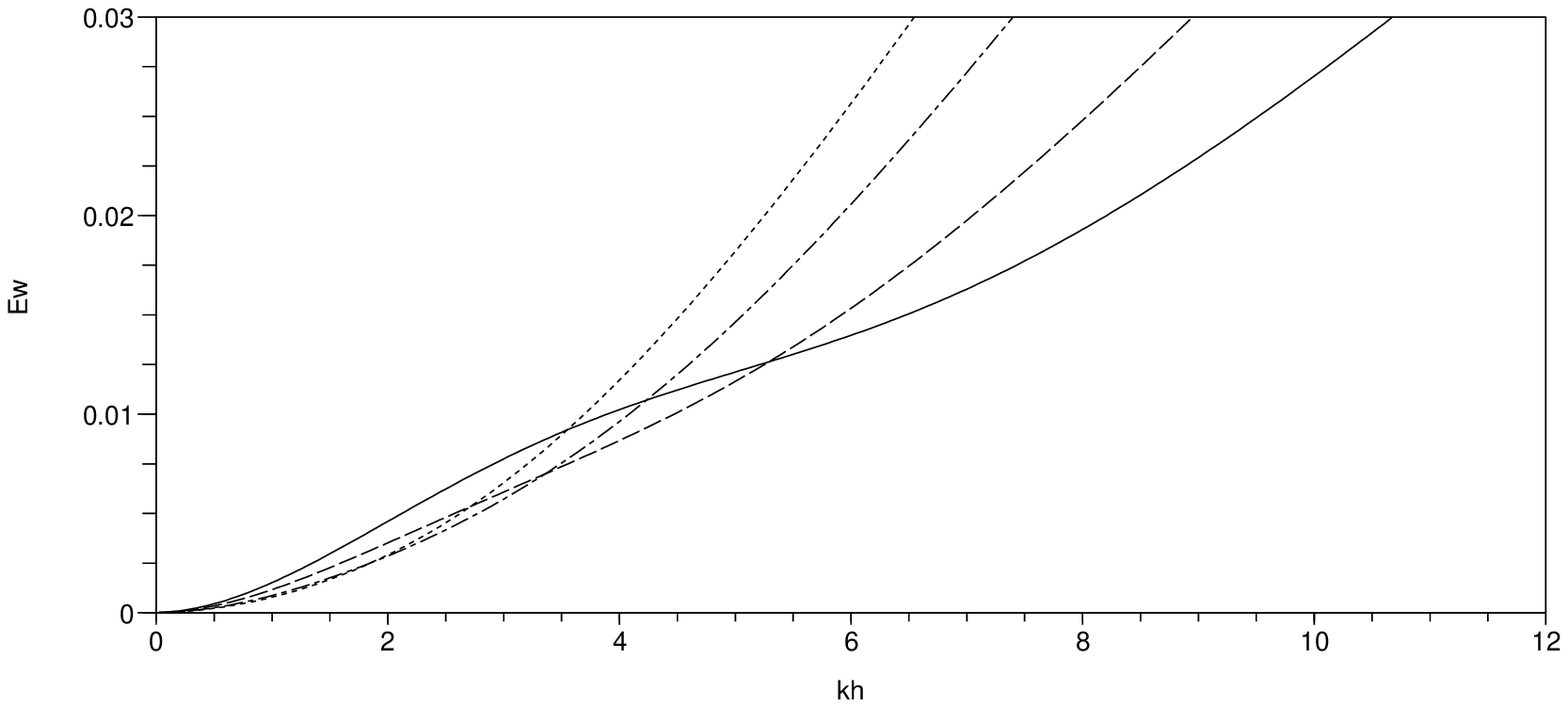,width=\linewidth}}
\caption{Depth-averaged errors on the vertical profile of the velocity potential (top) and vertical velocity (bottom) for $\sigma = 0.314$ (solid line), $\sigma = 0.365$ (dashed line), $\sigma = 0.428$ (dash-dotted line), and $\sigma = 0.473$ (dotted line).}
\label{figure3}
\end{figure}

Figure \ref{figure3} plots the depth-averaged errors $E_{\phi}$ (upper panel) and $E_{w}$ (lower panel) on the vertical profiles of the velocity potential and the vertical velocity. We remark that the difference between the errors obtained with each value of $\sigma_{opt}$ remains very small in shallow water. On the contrary, the benefit for taking $\sigma=0.314$ clearly appears for both the velocity potential and the vertical velocity in deep water: for the vertical velocity profile, a $1\%$ error is reached at $kh \approx 4$ and a $2\%$ error at $kh \approx 8$. As far as the velocity potential is concerned, a $1\%$ error is reached at $k \approx 6.5$ and a $2\%$ error is reached at $kh \approx 10$. By comparison, the model derived by Madsen \textit{et al.} (2002, 2003) yields a $2\%$ error at $kh = 12$ for both horizontal and vertical velocity profiles. The difference between the errors on the velocity potential and the vertical velocity component can be attributed to the fact that we have neglected the fourth-order derivative term in the expression (\ref{padéé}) of $w_i(z)$. As mentioned earlier, this choice entails to sacrifice some accuracy on the profile of the vertical velocity. Nevertheless, the global accuracy for both vertical profiles is still very good, up to the deep water value $kh = 10$.

\noindent This analysis of the phase celerity and velocity profile errors does not exhibit any major advantage for the choice $\sigma=0.473$ instead of $\sigma=0.314$ in shallow water. The additional errors made with $\sigma=0.314$ are at most of order $0.2\%$ in shallow and intermediate water. On the other hand, the advantage of this value clearly appears in deep water as it appreciably extends the linear range of validity of the model, especially for the vertical profiles. Therefore, we decide to adopt the value $\sigma=0.314$ in the sequel.

\subsection{Improved model with tightened shoaling properties}

\begin{figure}
\psfrag{Shoaling gradient}{{\footnotesize Shoaling gradient}}
\psfrag{kh}[][][0.8][0]{$kh$}
\center{\epsfig{file=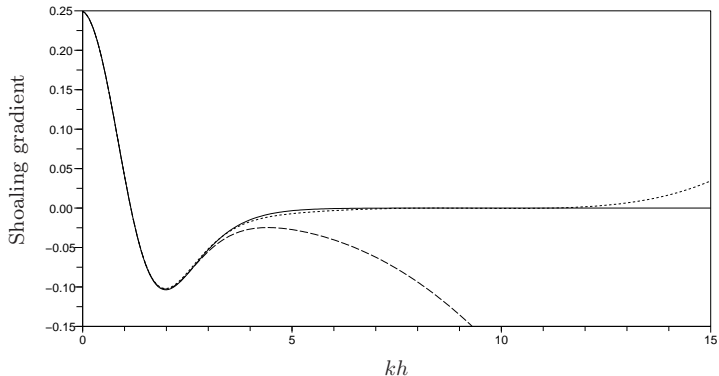,width=0.88\linewidth}}
\caption{Shoaling gradient: model with ($r = 0.0076$; dotted line) and without ($r=0$; dashed line) optimization with $\sigma = 0.314$, and exact shoaling gradient (solid line).}
\label{figure4}
\end{figure}

\noindent We now consider the linear optimization of the shoaling gradient properties. As shown on figure \ref{figure4}, the model linear shoaling gradient $\gamma_0$ (dashed line) only matches the exact linear shoaling gradient $\gamma_s$ (solid line) up to $kh = 3$ and swiftly departs from it beyond that value. We have attempted to optimize this shoaling gradient individually, but no value of $\sigma$ makes the two curves fit beyond $kh = 3$.

Therefore, a more subtle optimization is needed. This can be achieved using the following method. Going back to the full formulation of the approximate static Dirichlet-Neumann operator $\mathcal{G}_0^{app}[h]$ (\ref{TDN}), we introduce a new constant parameter $r$ and apply the differential operator $1 + r h \nabla h \cdot \nabla$ to the last equation of (\ref{TDN}), which yields 
\begin{equation}\label{prec}
\begin{array}{lll}
\vspace{0.25em}
P_0^{\ast} w_0 & \hspace{-0.5em}\ms & \hspace{-0.5em}\Big[ - {\beta_1^{\ast}} \Delta - r h \beta_1^{\ast} \nabla h \cdot \nabla\Delta \Big] \phi_1^{\ast}\\ 
& & \hspace{-0.25em}+ \Big[ 1 - {\alpha_1^{\ast}} \Delta - ({\varepsilon_1^{\ast}} - r h) \nabla h \cdot \nabla - r h \alpha_1^{\ast} \nabla h \cdot \nabla\Delta \Big] w_1^{\ast}\;,
\end{array}
\end{equation}
where $\displaystyle{P_0^{\ast} = 1 + \left(\frac{\sigma}{2}\beta_1^{\ast} + r h\right) \nabla h \cdot \nabla}$.\\
This new formulation does not allow further optimization yet, since the additional terms cancel out in the shoaling analysis. However, an interesting option is to neglect the third-order derivative on $w_1^{\ast}$ in the slope terms. By doing this, we greatly improve the shoaling gradient properties, as we will see in the sequel. For the moment, we rewrite (\ref{prec}) as follows:
\begin{equation}\label{wazaatmp}
P_0^{\ast} w_0 = \Big[ - {\beta_1^{\ast}} \Delta - r h \beta_1^{\ast} \nabla h \cdot \nabla\Delta \Big] \phi_1^{\ast} + \Big[ 1 - {\alpha_1^{\ast}} \Delta - ({\varepsilon_1^{\ast}} - r h) \nabla h \cdot \nabla \Big] w_1^{\ast}\;.
\end{equation}
We then take advantage of (\ref{astuce}) to eliminate the third-order derivative from the previous equation, and obtain
\begin{equation}\label{wazaa}
\begin{array}{lll}
\vspace{0.25em}
P_0^{\ast} w_0 & \hspace{-0.5em}= & \hspace{-0.5em}\Big[ - \beta_1^{\ast} \Delta - \frac{6 r}{\sigma} \nabla h \cdot \nabla \Big] \phi_1^{\ast} + \Big[ 1 - {\alpha_1^{\ast}} \Delta - ({\varepsilon_1^{\ast}} + 2 r h) \nabla h \cdot \nabla \Big] w_1^{\ast} \\
& & \hspace{-0.25em}+ \Big[ \frac{6 r}{\sigma} \nabla h \cdot \nabla \Big] \phi_0\;.
\end{array}
\end{equation}
This implies to redefine the approximate operator $\mathcal{G}_0^{app}[h]$ as follows:
\begin{equation}\label{G0app3}
\mathcal{G}_0^{app\,}[h]  =
Q_0^{\ast}
\left(
\begin{array}{cccc}
\mathcal{N}_1 & 0 & 0 & 0\\
0& \mathcal{N}_2 & 0 & 0\\
0 & 0 & \mathcal{N}_3 & 0\\
0 & 0 & 0 & 0
\end{array}
\right)
\mathcal{R}
\left(\begin{array}{c}P_0\\Q_1\\Q_2\\Q_3\end{array}\right)\,.
\end{equation}
where
\begin{equation}\label{G0app4}
\left\{
\begin{array}{l}
\mathcal{N}_1 = - \beta_1^{\ast} \Delta - \frac{6 r}{\sigma} \nabla h \cdot \nabla\,,\;\;
\mathcal{N}_2 = 1 - {\alpha_1^{\ast}} \Delta - ({\varepsilon_1^{\ast}} + 2 r h) \nabla h \cdot \nabla\,,\\
\mathcal{N}_3 = \frac{6 r}{\sigma} \nabla h \cdot \nabla\,,\;\;
Q_0^{\ast} = 1 - \left(\frac{\sigma}{2}\beta_1^{\ast} + r h\right) \nabla h \cdot \nabla \ms (P_0^{\ast})^{-1}\,.
\end{array}
\right.
\end{equation}
Our modified model is thus identical to (\ref{model}) but with $\mathcal{G}_0^{app\,}[h]$ as redefined above. Starting from it, the new linearized model remains essentially the same, except that the second equation of (\ref{linf}) now is
$$
\partial_t N^{\ast} + \Big[ \beta_1^{\ast} \partial_x^2 + \frac{6 r}{\sigma} h_x \partial_x \Big] \phi_1^{\ast} - \Big[ 1 - \alpha_1^{\ast} \partial_x^2 - ({\varepsilon_1^{\ast}} + 2 r h) h_x \partial_x \Big] w_1^{\ast} - \Big[ \frac{6 r}{\sigma} h_x \partial_x \Big] \phi^0 \ms 0\,,
$$
with $N^{\ast} = P_0^{\ast} \eta$. This new formulation does not modify the phase celerity and the vertical profiles, but allows to further minimize the shoaling gradient error. We can now compute the new shoaling gradient $\gamma_0$. Optimizing the parameter $r$ so that the error $\mathcal{E}_{\gamma}(K,\sigma)$ is minimized for $K = 10$ and $\sigma = 0.314$ yields $r = 0.0076$.

\begin{rmrk}
A dimensional analysis shows that this small value of $r$ is coherent with our earlier choice to neglect the third-order derivatives of $w_1^{\ast}$ in (\ref{wazaatmp}).
\end{rmrk}

\noindent Figure \ref{figure4} displays the optimized shoaling gradient for the previous value for $r$. The improvement is quite impressive since the new shoaling gradient exhibits a very good agreement up to $kh \approx 12$. This accuracy is the same as that reached by the model of Madsen \textit{et al.} (2002, 2003).

\vspace{1em}

\noindent To sum up, our final double-layer Boussinesq-type model consists of (\ref{model}) and (\ref{G0app3})--(\ref{G0app4}) with $\sigma = 0.314$ and $r=0.0076$. This model exhibits excellent linear properties: the phase celerity is accurate up to $kh = 28$, the vertical velocity profiles are accurate up to $kh = 10$ for the velocity potential and up to $kh = 8$ for the vertical velocity component, and the shoaling gradient is accurate up to $kh = 12$. We emphasize that these properties are not affected by a slight variation of $\sigma$, which makes the model robust towards the parameter $\sigma$. These results are quite similar to those obtained by Madsen \textit{et al.} (2002, 2003), but the main advantage of the present model is that it contains lower-order derivatives and fewer equations, especially in 2DH. The present double-layer approach is hence a very good alternative to the most advanced high-order Boussinesq models as it offers almost the same linear properties with a lower complexity.

\section{Numerical simulations: nonlinear behaviour}

On the basis of the model (\ref{model}) derived in \S 2 (we do not use here the version derived in \S 4$\,f$ since we consider a flat bottom), a classical finite-difference scheme is developed to study numerically some nonlinear properties of the model in 1DH.

We consider the propagation of two-dimensional periodic and regular nonlinear waves, without change of form, over a flat bottom and without any ambient flow field. For this situation, numerical reference solutions can be obtained by the so-called \textit{stream function method}, or more precisely the Fourier approximation of the stream function (Dean 1965; Rienecker \& Fenton 1981). Unlike analytical wave theories (such as Stokes or cnoidal wave theories), this numerical approach is applicable whatever the relative water depth and steepness are, and very accurate solutions can be obtained by increasing the number of terms in the Fourier series (e.g. 10, 20, or 50 if necessary for very steep waves). This method was previously implemented in a software called Stream\_HT by one of the authors (Benoit \textit{et al.} 2002). For the selected application, the domain of interest covers one wave-length ($L = 64$\3m ; $k = 2\pi/L = 0.098$\3rad\3m$^{-1}$) and periodic conditions are imposed at the two lateral boundaries. We consider a still water depth of $h = 96$\3m, so that the relative water depth is $h/L = 3/2$ yielding $kh = 3\pi \approx 9.425$, which corresponds to deep water conditions. The wave height is chosen as $H = 6.4$\3m, so that the steepness is $H/L = 0.1$ or $kH/2 = \pi/10 \approx 0.314$, i.e. about $70\%$ of the theoretical maximum value of the steepness for a stable wave (Williams 1981). These conditions correspond to highly dispersive and very nonlinear waves. For this case, the period computed with the stream function approach (at order 20) is $T = 6.094$\3s, yielding a wave celerity of $C = L/T \approx 10.502$\3m\3s$^{-1}$. The solution obtained with the stream function approach for the free surface elevation and the free surface potential is imposed as the initial condition in our simulations.

The time integration scheme is a classical fourth-order four-stage
explicit Runge-Kutta scheme, which is known to possess a wide stability
region. However, owing to the nonlinear nature of the considered test
case, this scheme can develop some high-frequency instabilities. To
avoid such instabilities, an 8th-order Savitsky-Golay smoothing filter
is applied twice after each time step to $\eta$, $\widetilde{\phi_1}$,
and $\widetilde{w_1}$. The price to pay for this filter is a negligible
loss of accuracy of the model. As far as spatial discretization is
concerned, all derivative operators in (\ref{model}) are replaced by
centered fourth-order finite difference approximations combined with
periodic boundary conditions. The covered domain (of one wave-length) is
discretized with 32 cells of constant size (equal to $2$\3m) and a
time-step of $0.122$\3s (corresponding to $T/50$) is used during the
simulations. \textcolor{black}{We have verified that using a refined mesh of 128 cells and 200 time steps per period does not yield any significant improvement.}

\begin{figure}
\center{
\psfrag{etaenm}[bc][tl]{${\scriptstyle\eta(m)}$}
\psfrag{xenm}[t][b]{${\scriptstyle x(m)}$\;\;}
\psfrag{Free surface at t = 10T}[Bc][Bl][1][0]{\;\;\;\;\;{\scriptsize Free surface elevation at $t=10\3T$}}
\epsfig{file=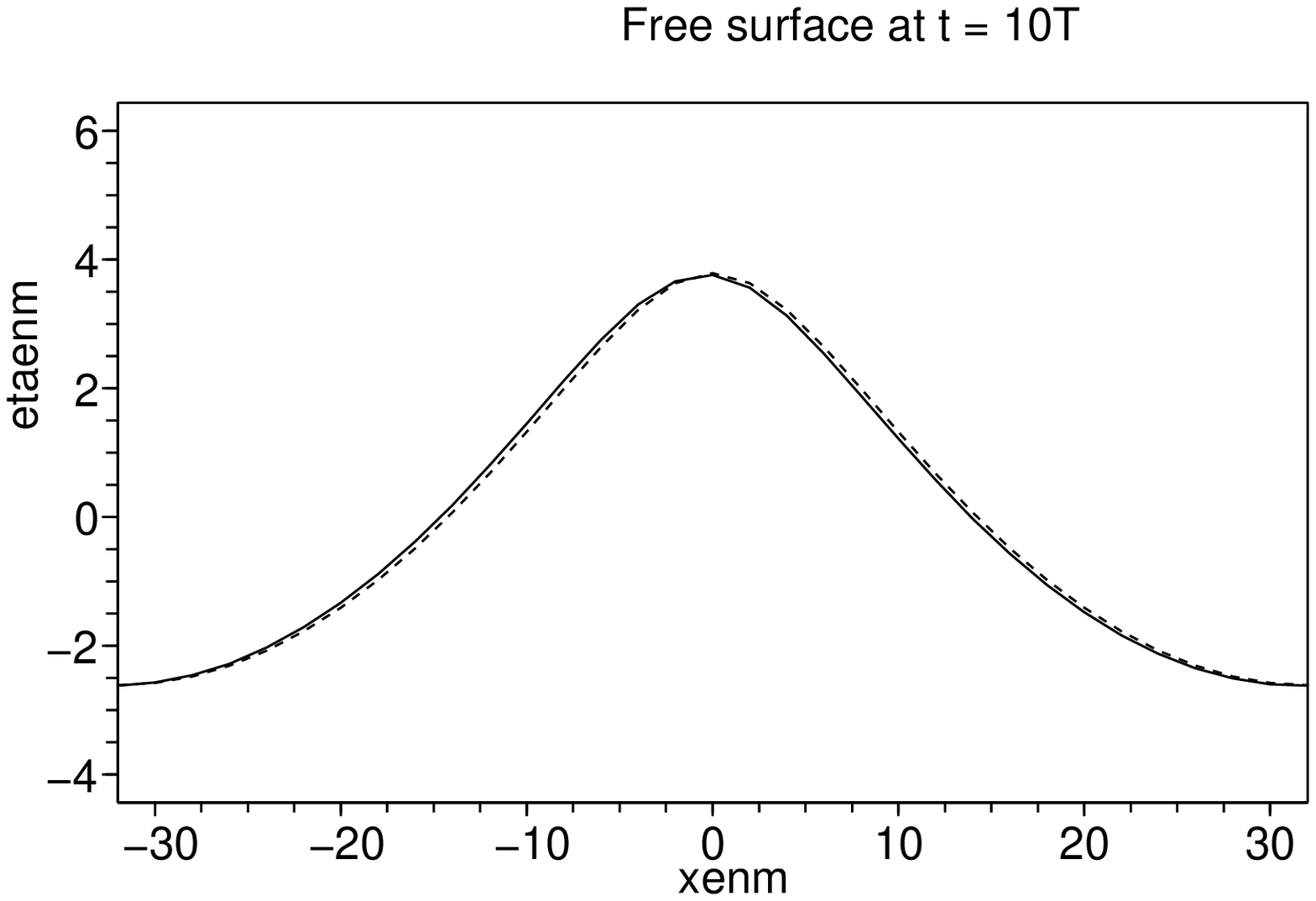,width=6cm}
\psfrag{etaenm}[bc][tl]{${\scriptstyle\eta(m)}$}
\psfrag{xenm}[t][b]{${\scriptstyle x(m)}$\;\;}
\psfrag{Free surface at t = 25T}[Bc][Bl][1][0]{\;\;\;\;\;\,{\scriptsize Free surface elevation at $t=25\3T$}}
\epsfig{file=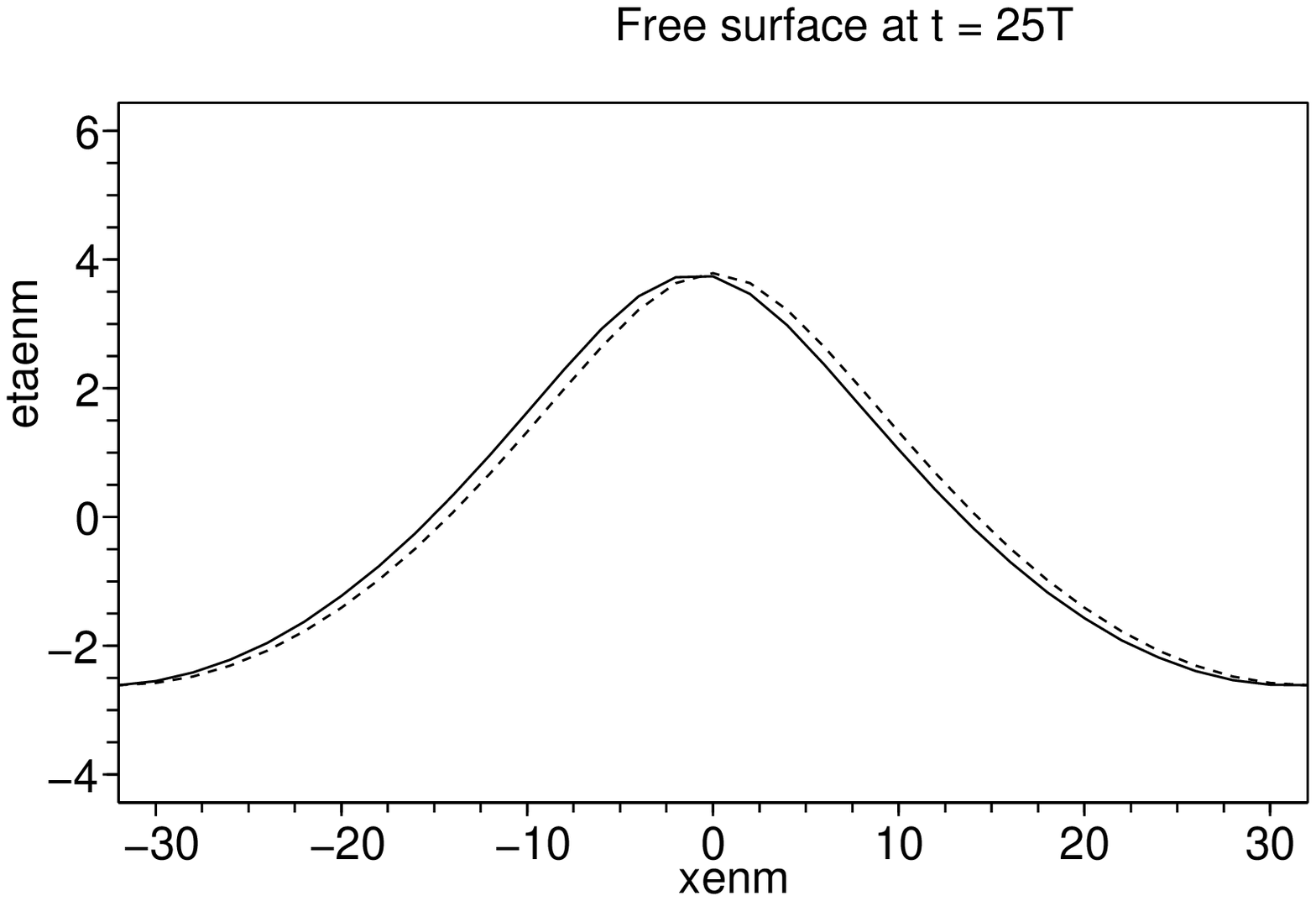,width=6cm}
}
\center{
\psfrag{phienmdeuxsmoinsun}[bc][tl]{${\scriptstyle\widetilde{\phi_1}(m^2\text{\hspace{-0.25em}/s})}$}
\psfrag{xenm}[t][b]{${\scriptstyle x(m)}$\;\;}
\psfrag{Velocity potential at the free surface at t = 10T}[Bc][Bl][1][0]{\hspace{3.25em}{\scriptsize Free surface potential at $t=10\3T$}}
\epsfig{file=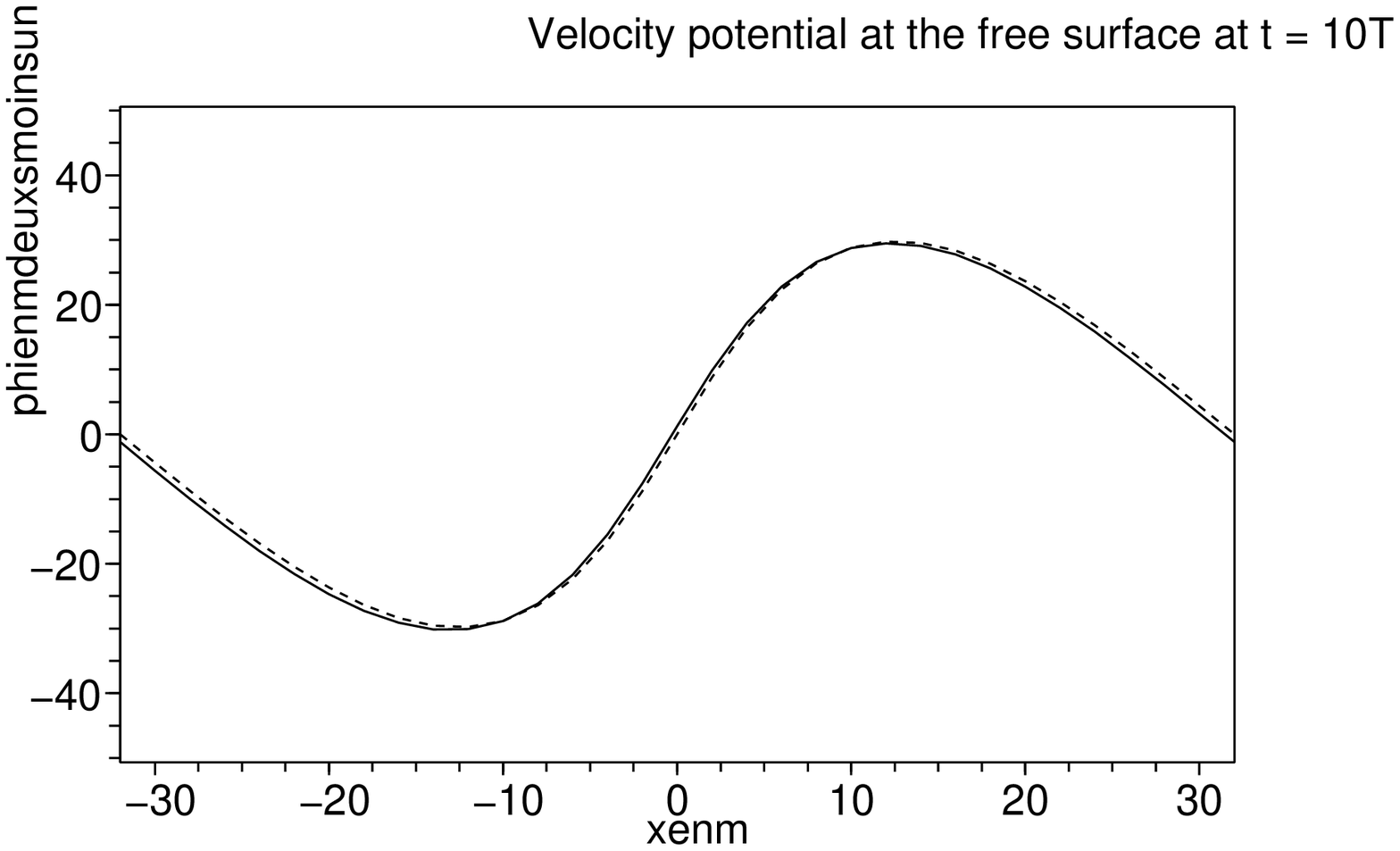,width=6cm}
\psfrag{phienmdeuxsmoinsun}[bc][tl]{${\scriptstyle\widetilde{\phi_1}(m^2\text{\hspace{-0.25em}/s})}$}
\psfrag{xenm}[t][b]{${\scriptstyle x(m)}$\;\;}
\psfrag{Velocity potential at the free surface at t = 25T}[Bc][Bl][1][0]{\hspace{3.25em}{\scriptsize Free surface potential at $t=25\3T$}}
\epsfig{file=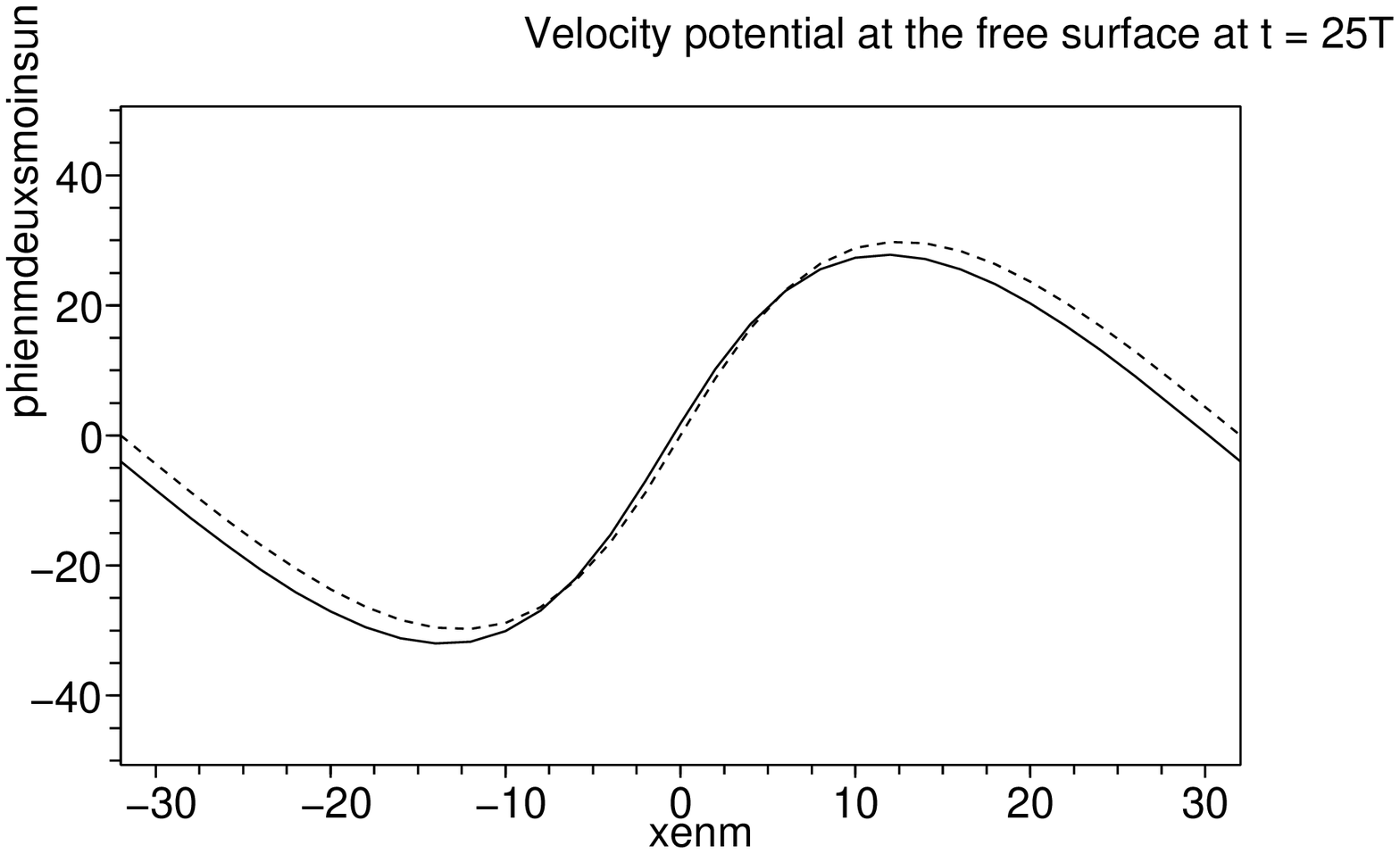,width=6cm}
}
\caption{Snapshots of the free surface elevation (top) and the free surface velocity potential (bottom) at $t=10\3T$ (left) and $t=25\3T$ (right). The solid line is our model for $\sigma = 0.314$ and the dotted line is the reference solution computed using the stream function approach.}
\label{figure5}
\end{figure}
Numerical integration of the double-layer model (\ref{model}) is
performed over a duration of $25\3T$. Simulations have been performed
with the deep-water ($kh=10$) optimal value $\sigma = 0.314$\textcolor{black}{, even if any value in $[0.28,0.36]$ would lead to very
  similar results. Outside this range, the results quickly
  deteriorate}. Results obtained after durations of $10\3T$ and $25\3T$
are plotted on figure \ref{figure5} for the free surface elevation and
the free surface velocity potential. The results are compared to the
reference solution obtained with the stream function approach (which
propagates at constant speed and without change of form). The results
appear to be very good since the two curves fit very well until
$t\approx20\3T$, after which small discrepancies become
observable. \textcolor{black}{Since grid convergence has been verified,
  this difference can be attributed to the approximation in (\ref{closure}), where we have neglected fourth-order (and higher) nonlinear terms.} However, we remark that the global aspect of the model curve still corresponds to that of the reference solution at $t=25\3T$. Furthermore, the free surface elevation computed with our model only shows a phase shift error with the reference solution: the forms are the same and the amplitude of the waves are equal. An interesting remark is that we can use these results to compute the nonlinear phase celerity error approximatively. Indeed, taking for instance the free surface elevation results and measuring the distance between the crests of the two curves yields an approximate value of the difference of celerity between these curves. This value provides us a measure of the nonlinear celerity error of the model. We found that our model with $\sigma_{opt} = 0.314$ exhibits a nonlinear phase celerity error of about $0.08\%$, which is an impressive result. To conclude, the model shows an excellent nonlinear behaviour, and we can expect its nonlinear range of validity to reach up to $kh = 10$, at least for flat bottom conditions.

\label{lastpage}

\end{document}